\documentclass[prc,aps,a4paper,groupedaddress,superscriptaddress,nofootinbib
,preprintnumbers,twocolumn]{revtex4}
\usepackage{graphicx}
\usepackage{amsfonts}
\usepackage{amssymb}
\usepackage{amsmath}
\usepackage{natbib}
\usepackage{dcolumn}
\usepackage{bm} 
\usepackage{color,xcolor}
\usepackage{multirow}
\usepackage{caption}
\usepackage{hyperref}
\usepackage[normalem]{ulem}
\usepackage[export]{adjustbox}
\usepackage{float}

\newcommand{\bwt}{\begin{widetext}}
	\newcommand{\ewt}{\end{widetext}}
\newcommand{\beq}{\begin{equation}}
	\newcommand{\eeq}{\end{equation}}
\newcommand{\bea}{\begin{eqnarray}}
	\newcommand{\eea}{\end{eqnarray}}

\begin{document}

\title{ 
General predictions of neutron star properties using unified relativistic mean-field equations of state 
}
\author{Luigi Scurto}
\email{lscurto@student.uc.pt}
\affiliation{CFisUC, Department of Physics, University of Coimbra, 3004-516 Coimbra, Portugal}
\author{Helena Pais} 
\email{hpais@uc.pt} 
\affiliation{CFisUC, Department of Physics, University of Coimbra, 3004-516 Coimbra, Portugal} 
\author{Francesca Gulminelli}
\email{gulminelli@lpccaen.in2p3.fr}
\affiliation{Normandie Univ., ENSICAEN, UNICAEN, CNRS/IN2P3, LPC Caen, F-14000 Caen, France}

\begin{abstract}

In this work we present general predictions for the static observables of neutron stars (NSs) under the hypothesis of a purely nucleonic composition of the ultra-dense baryonic matter, using Bayesian inference on a very large parameter space conditioned by both astrophysical and nuclear physics constraints. 

The equation of states are obtained using a unified approach of the NS core and inner crust within a fully covariant treatment based on a relativistic mean-field Lagrangian density with density dependent couplings. The posterior distributions are well compatible with the ones obtained by semi-agnostic meta-modelling techniques based on non-relativistic functionals, that span a similar portion of the parameter space in terms of nuclear matter parameters, and we confirm that the hypothesis of a purely nucleonic composition is compatible with all the present observations.

We additionally show that present observations do not exclude the existence of very massive neutron stars with mass compatible with the lighter partner of the gravitational event GW190814 measured by the LIGO-Virgo collaboration. 
Some selected representative models, that respect well all the constraints taken into account in this study, and approximately cover the residual uncertainty in our posterior distributions, will be uploaded in the \textsc{CompOSE}  database for use by the community. 
 
\end{abstract}
\maketitle

\section{Introduction}

The analysis of multi-messenger data on compact stars require models for the equation of state (EoS) of dense matter that are sufficiently flexible to account for the present uncertainties of the theoretical modelling, thus avoiding artificial bias in the parameter estimation, see refs.\cite{Oertel:2016bki,Baym_2018} for recent reviews on the EoS modelling. 

A paramount - though not unique - example of this statement is given by the estimation of neutron star radii from gravitational wave (GW) signals observed by ground based interferometers from the merging of compact binaries \cite{ Abbott:2018exr}. Indeed, given the mass of a neutron star, its tidal deformability under the gravitational field of the companion in the late inspiral dynamics is one-to-one correlated to the star radius, for a given equation of state model. Even if the leading order adiabatic tidal effects enter the phase of the waveform only at the fifth post-Newtonian order\cite{Hinderer08,Hinderer09} and can therefore be extracted only from loud signals with a a high signal-to-noise ratio, important constraints on the tidal parameter were already obtained by the LVK collaboration, particularly from the famous GW170817 event \cite{TheLIGOScientific:2017qsa}. New observations from the ongoing O4 and upcoming O5 run, and next generation detectors such as Einstein Telescope  Cosmic Explorer are expected to further tighten the tidal deformability constraint through more precise and more numerous observations \cite{Maggiore:2019uih}. 
A precise estimation of the correlation between the tidal polarizability parameter $\Lambda$ and the star mass $M$ can therefore lead to radius evaluations that can be more precise than  estimations from X-ray bursts or pulse timing measurements \cite{Ozel:2016oaf,Steiner:2017vmg,Bogdanov_2019,Bogdanov_2021}, at the same time providing precious information on the behavior of ultra-dense matter in a regime that is completely inaccessible to laboratory experiments.

The most model-independent approach in this respect is given by agnostic (parametric or non-parametric) EoS modelling, such as piecewise polytropes \cite{Read:2008iy}, spectral parameterization \cite{Lindblom:2010bb,Lindblom:2012zi,Lindblom:2013kra}, Gaussian process-based sampling \cite{Landry:2018prl,Essick:2019ldf,Landry:2020vaw,Legred_2022}, or neural networks\cite{Han2023} that do not impose a specific form of the EoSs and are only restricted by the requirement of causality and thermodynamic stability. Although very
powerful for EoS inference from astrophysical data, the common drawback of these agnostic models is that they do not provide information on the internal composition of neutron stars, and more generally on the properties of the strong interaction in the non-perturbative domain of QCD. For this reason, semi-agnostic models have been proposed \cite{Margueron:2017eqc,Margueron:2017lup,Zhang_2018,LimHolt,PhysRevC.102.045808,Malik:2018zcf,Providencia:2023rxc}, that do not directly model the functional relation between pressure and energy density, but infer this quantity by imposing the equilibrium of weak interactions to a flexible parametrisation of the energy density of an electrically neutral system composed of leptons and baryons. In this meta-modelling approach, the parameters of the energy functional are chosen such as to explore the full uncertainty domain of microscopic models of dense matter, with minimal hypotheses on the relevant degrees of freedom.
The simplest realization of this approach consists in considering only nucleonic (neutron and proton) degrees of freedom for the baryonic part, and express the interaction part of the energy density as a polynomial expansion around the equilibrium density of symmetric nuclear matter, where the behavior is best constrained by laboratory experiments \cite{Kortelainen2010,Brown2013,DANIELEWICZ2017147,Khan2013,RocaMaza2016}. Alternatively, expansions around pure neutron matter have been proposed, where ab-initio low-energy nuclear theory can provide powerful constraints \cite{Gezerlis_2010,Hebeler_2013,Tews_2013,Lynn_2016,Tews_2017,Drischler_2019,Drischler:2021kxf}. Combined with Bayesian statistical analysis, this meta-modelling technique can be used to assess the compatibility of specific dense matter theories to the existing nuclear and astrophysical data, and also as a null hypothesis to infer from the astrophysical data the presence of exotic non-nucleonic degrees of freedom \cite{universe7100373,Pfaff:2021kse,Iacovelli_2023,Mondal:2023gbf}. Finally, a complete consistency is insured between the behavior of the homogeneous star core, that contains the highest uncertainties, and the one of the solid crust, that is well constrained by nuclear theory.   

The main drawback of this approach is that polynomial expansions by construction lead to unphysical acausal behaviors at high density. Even if causality can be restored by explicitly selecting models that become superluminal at densities overcoming the central density of the highest mass $M_{max}$ star allowed by the hydrostatic equilibrium Tolmann-
Oppenheimer-Volkof (TOV) equation, the average value of the speed of sound in the center of the star typically exceeds 80\% of the speed of light \cite{Pfaff:2021kse}, calling for a relativistic treatment of the energy functional\cite{Malik:2018zcf,Malik_2022,Malik:2023mnx}.
In a recent work, P.Char and collaborators \cite{Char_2023} have proposed a relativistic version of the meta-modelling technique, where not only leptons, but also baryons are treated within a covariant Lagrangian formulation based on the relativistic mean-field (RMF) theory. To realistically capture the theoretical uncertainty of the interaction at high density, density dependent couplings were used, with the functional form proposed in ref.\cite{Gogelein:2007qa}, that will be noted GDFM in the following after the authors' names. The expression of the couplings comprises a large number of parameters both in the isoscalar and in the isovector sector, such as to cover all the possible behaviors of the poorly known density dependence of the symmetry energy, and therefore a-priori allow for a very large range of possible compositions in dense matter.  

In this study we continue the endeavour of ref.\cite{Char_2023} and study in greater detail the effect of astrophysical and nuclear physics constraints and the different likelihood models on the posterior distribution of astrophysical observables and  crustal properties of NS. 
At variance with ref.\cite{Char_2023} and other previous studies along the same lines \cite{Malik_2022,Malik:2023mnx}, that use a different expression for the density-dependent couplings involving a smaller set of parameters,  in this work we build both the core and the inner crust of the star, employing the density-dependent GFDM functional form, as in Ref.~\cite{Char_2023}, and using a compressible liquid drop (CLD) approximation \cite{BAYM1971225,DouchinHaensel,Newton_2013,Pais_2015,Carreau:2019zdy,DinhThi2023} for the ions in the inner crust with a surface tension optimized on experimentally measured nuclear masses\cite{Carreau:2019zdy}.

  We divide our analysis in three main parts.
  In the first part we study the effect of astrophysical constraints, in particular the constraint on the radius of NS coming from the X-ray observations of the NICER observatory\cite{Riley:2019yda,Miller:2019cac,Riley:2021pdl,Miller:2021qha}.
  In the second part we study the effect of the constraints coming from chiral effective field theory  ($\chi$-EFT) \cite{Hebeler_2013,Tews_2013,Lynn_2016,Drischler_2019,Drischler:2021kxf}, observing how different prescriptions for the application of this constraint affect the posterior distribution. Finally in the third part we focus on the effect of experimental constraints on the properties of nuclear matter. In particular we discuss the compatibility of the constraints extracted from the parity violating electron scattering PREX and CREX experiments\cite{PREX:2021umo,CREX:2022kgg,Reed:2023cap} with the parameter space allowed by the behavior of the GDFM Lagrangian for the different nuclear matter paramerters (NMPs).  
  
  The structure of the paper is as follows. In the following section we present the theoretical framework used for the construction of the EoS used in our study. In Section \ref{Sec_Bayes} we present the procedure used for the construction of our prior distribution and for the estimate of the weight assigned to each model for the evaluation of our posterior distributions. In Section \ref{Sec_res} we present our results for the three parts of our study and finally in Section \ref{Sec_fin} we draw some conclusions.

\section{Construction of Neutron Star EoS}\label{Sec_theory}

\subsection{Homogeneous Matter}

The EoSs used in this study are built within a RMF approximation, where the interaction between the nucleons is mediated by three types of mesons: the isoscalar-scalar meson $\sigma$ of mass $m_\sigma$, the isoscalar-vector meson $\omega$ of mass $m_\omega$, and the isovector-vector meson $\rho$ of mass $m_\rho$. Following ref.\cite{Char_2023}, we do not include the $\delta$ meson, i.e. a scalar isovector channel. 
The inclusion of this channel in future works might be useful to access more complex, non-monotonic behaviors of the density dependence of the symmetry energy \cite{Li:2022okx}. However, we will show that a large variety of symmetry energy behaviors can be modelled by the simple vector coupling, thanks to the complex behavior of the density dependence employed.

The Lagrangian density of our models is given by the standard form:
\begin{equation}
\mathcal{L}=\sum_{i=p,n}\mathcal{L}_i+\mathcal{L}_L+\mathcal{L}_\sigma+\mathcal{L}_\omega+\mathcal{L}_\rho \, .
\end{equation}
Here, $\mathcal{L}_L$ is the Lagrangian of the lepton component, including electrons and muons and treated as a free fermion gas, while the baryon contribution is given by: 
\begin{equation}
        \mathcal{L}_i=\bar{\psi}_i\big[\gamma_\mu  i\partial^\mu-g_\omega V^\mu-\frac{g_\rho}{2}\mathbf{\tau}\cdot\mathbf{b}^\mu-M^*\big]\psi_i \, ,
\end{equation}
where  $M^*_i= M^*=M-g_\sigma\phi$ is the effective nucleon mass that is taken as isospin independent, with $M$ being the bare nucleon mass, $M=M_n=M_p=938.9$ MeV 
, while 
the mesonic components  are given by 
\begin{eqnarray}
    \mathcal{L}_\sigma&=&\frac{1}{2}\bigg(\partial_\mu\phi\partial^\mu\phi-m_\sigma^2\phi^2\bigg) \, ,\\
    \mathcal{L}_\omega&=&-\frac{1}{4}\Omega_{\mu\nu}\Omega^{\mu\nu}+\frac{1}{2}m_\omega^2V_\mu V^\mu\, ,\\
    \mathcal{L}_\rho&=&-\frac{1}{4}\mathbf{B}_{\mu\nu}\cdot\mathbf{B}^{\mu\nu}+\frac{1}{2}m_\rho^2\mathbf{b}_\mu\cdot \mathbf{b}^\mu \, ,
\end{eqnarray}
with the tensors written as
\begin{eqnarray}
\Omega_{\mu\nu}&=&\partial_\mu V_\nu - \partial_\nu V_\mu \, , \\
\mathbf{B}_{\mu\nu}&=&\partial_\mu \mathbf{b}_\nu - \partial_\nu \mathbf{b}_\mu - g_\rho \left(\mathbf{b}_\mu \times \mathbf{b}_\nu \right)\, .
\end{eqnarray}

The leptonic component of the Lagrangian is given by 
\begin{equation}
\mathcal{L}_L=\sum_{i=e,\mu} \bar{\psi}_i\big[\gamma_\mu i \partial^\mu-m_e\big]\psi_i,
\end{equation}
where the sum takes into account the contribution of both electrons and muons.  

In the mean-field approximation, the meson fields are static and constant, and the energy density at a baryon density $\rho$ and asymmetry $\rho_3=\rho_p-\rho_n$ takes the simple form 
\begin{eqnarray}
\mathcal{E}(\rho,\rho_3)&=& \sum_{i=n,p,e,\mu} \langle \bar{\psi}_i\gamma_0 k_0{\psi}_i\rangle - \langle \mathcal{L} \rangle   \,   \\
&=& \sum_{i=n,p,e,\mu}   \mathcal{E}_{kin,i} 
  - \mathcal{E}_{field}   \, , \label{En_dens_hom}
\end{eqnarray}
with
\begin{eqnarray}
   \mathcal{E}_{kin,i}&=& \frac{2J_i+1}{2\pi^2}\int_0^{k_{F,i}} dk\,k^2 \sqrt{k^2+M_i^{*2}} \\
  \mathcal{E}_{field}&=&\frac 12 m_\omega^2 \omega_0^2 + \frac 12 m_\rho^2 b_{3,0}^2 - \frac 12 m_\sigma^2 \phi_0^2  \nonumber\\
  &-&   \sum_{i=n,p} \left ( g_\omega\omega_0\rho_i\pm g_\rho b_{3,0} \rho_i \right ) .
\end{eqnarray}
Here, leptons are given their free mass ($M^*_e=m_e, M^*_\mu=m_\mu$), $J_i$ and $k_{F,i}$ are ground state spins and Fermi momenta, respectively, with $\rho_{i}=k_{F,i}^3/(3\pi^2)$,
and the subscripts $0$ note the expectation values of the static fields ($\langle\phi\rangle=\phi_0$, $\langle V_\mu\rangle=\omega_0$, $\langle\mathbf{b}_\mu\rangle = b_{3,0} $), and the sign $+(-)$ corresponds to protons (neutrons).
The chemical potentials of the nucleons are given by
\begin{equation}    
\mu_i=\sqrt{k^2_{F,i}+M^{*2}}+g_\omega\omega_0\pm \frac{1}{2}g_\rho b_{3,0}+\Sigma_R 
\end{equation}
where $i=p,n$, 
and $\Sigma_R$ is the rearrangement term arising from the density dependence of the couplings:
\begin{align}
    \Sigma_R(\rho)=\frac{\partial g_\omega}{\partial\rho}\omega_0\rho+\frac{1}{2}\frac{\partial g_\rho}{\partial \rho}b_{0,3} \rho_3-\frac{\partial g_\sigma}{\partial \rho}\phi_0\rho_s,
\end{align}
$\rho_s$ being the scalar density :
\begin{align}
    \rho_s=\sum_{i=p,n} \langle \bar{\psi}_i{\psi}_i\rangle 
    =\frac{1}{\pi^2}\sum_{i=p,n} \int_0^{k_{F,i}} dk \,
    \frac{M^* k^2}{\sqrt{k^2+M^{*2}}} .
\end{align}
In the same mean-field approximation, leptons are an homogeneous free gas with chemical pontials simply given by
\begin{equation}
    \mu_i=\sqrt{k^2_{F,i}+m_i^{2}} \, ,
\end{equation}
with $i=e,\mu$. Finally, the pressure is given by

\begin{eqnarray}
P&=& \sum_{i=n,p,e,\mu} \mu_i \rho_i - \mathcal{E}\,   \\
&=& \sum_{i=n,p,e,\mu} P_{kin,i}
  -\frac 12 m_\omega^2 \omega_0^2 + \frac 12 m_\rho^2 b_{3,0}^2 - \frac 12 m_\sigma^2 \phi_0^2+\rho\Sigma_R  \nonumber \, . \label{En_dens_hom}
\end{eqnarray}
Where the "kinetic" contribution reads 
\begin{equation}
 P_{kin,i}=\frac{2J_i+1}{6\pi^2}\int_0^{k_{F,i}} dk\,\frac{k^4}{ \sqrt{k^2+M_i^{*2}}} .
\end{equation}

Charge neutrality $\rho_p=\rho_e+\rho_\mu$ and the weak equilibrium conditions complete the equations set for homogeneous core matter:
\begin{equation}
 \mu_n-\mu_p=\mu_e \; ; \; \mu_e=\mu_\mu   
\end{equation}

 For the couplings, we choose as in ref.\cite{Char_2023} the GDFM \cite{Gogelein:2007qa} ansatz, with each coupling in the form 
\begin{equation}
    g_i=a_i+(b_i+d_ix^3)e^{-c_i x}\, \label{eq:couplings}
\end{equation}
with $i=\sigma,\omega,\rho$, and $x=\rho/\rho_{sat}$, where $\rho_{sat}$ is the saturation density of the model. Each model is thus uniquely defined by 12 parameters, 4 for each meson coupling. This set of 12 parameters identifying each realization of the GDFM Lagrangian will be noted $\bf X=\{X_1,\dots,X_{12}\}$ in the following.

\subsection{Neutron Star Crust} \label{sec:crust}

All the EoS used in the study include an outer and an inner crust. For the outer crust we use the BSK22 model \cite{Pearson_2018}, while a unified inner crust is built for each model. In order to do so, we use the same GDFM model described in the previous section and a  CLD calculation \cite{Pais_2015,Scurto_2023} for the description of the heavy clusters in the crust.\\
In our realization of the CLD model, each Wigner-Seitz cell 
is composed of a constant electron background of energy density $\mathcal{E}_L$, a high-density ("cluster") part, labeled $I$, and a low-density ("gas") part, labeled $II$, both considered as homogeneous portions of nuclear matter at respective densities $\rho^I$ and $\rho^{II}$ .  The equilibrium proportion of cluster and gas and their composition are obtained by minimizing the total energy density of the cell, including the interface surface and Coulomb terms, that is given by
\begin{equation}
\mathcal{E} =f {\mathcal{E}}(\rho^I,\rho_I^{I})+(1-f) {\mathcal{E}}(\rho^{II},\rho_I^{II}) +\mathcal{E}_{Coul}+\mathcal{E}_{surf}+\mathcal{E}_L \, ,
\label{En_dens}
\end{equation}
with respect to four variables: the linear size of the cluster, $R_d$, the baryonic density $\rho^I$ and proton density $\rho_p^I$ of the high density phase, and its volume fraction  $f$. In this study, we only consider spherical clusters in order to slightly reduce the computational time, as it was shown by different authors \cite{Avancini_2008,Grill_2014,Pearson_2020,DinhThi2023} that the inclusion of exotic geometries ("pasta" phases) has a  negligible effect both on the EoS and on the crust-core transition point. In fact, in some calculations at $T = 0$ MeV and $\beta-$equilibrium matter, only the droplet configuration is present for the inner crust \cite{Avancini_2008}, and it was even argued in \cite{PhysRevC.105.015803} that shell effects could strongly suppress the non-spherical phases.
\\
{The Coulomb and surface terms are given by \cite{Ravenhall_1983}
\begin{eqnarray}
\mathcal{E}_{Coul}&=&2f e^2\pi\Phi R_d^2 \left(\rho_p^I-\rho_p^{II}\right)^2 \, , \label{eq:ecoul} \\
\mathcal{E}_{surf}&=&\frac{3f\sigma}{R_d} \label{eq:esurf}
\end{eqnarray}
where $\Phi$ is given by
\begin{equation}
\Phi=\frac{1}{5}\left(2-3f^{1-2/3}+f\right) \, 
\end{equation}
and where the surface tension $\sigma$ reads
\begin{equation}
    \sigma=\sigma_0\frac{b+2^{4}}{b+y_{p,I}^{-3}+(1-y_{p,I})^{-3}}
\end{equation}
where $y_{p,I}$ is the proton fraction of the dense phase and where the parameters $\sigma_0$ and $b$ are optimized for each model, through a fit over the measurements of the nuclear masses\cite{Wang_2017}, as previously done in \cite{Carreau:2019zdy,DinhThi2023,DinhThi2023a}. This procedure ensures an estimation of the surface tension that is consistent with each specific model, without increasing the size of the parameter space. \\
The density and pressure at the crust-core transition are consistently obtained for each realization of the GDFM model by comparing, for each baryonic density $\rho$, the energy density of inhomogeneous matter Eq.(\ref{En_dens}) to the energy density of homogeneous matter with a composition determined by the condition of $\beta$-equilibrium.
Since the very same Lagrangian density is used to evaluate the energy densities appearing in Eq.(\ref{En_dens_hom}) and (\ref{En_dens}), if stringent constraints are applied at low baryonic density, such as the ab-initio neutron matter calculations from many-body perturbation theory using chiral interactions, they will effectively affect also the high density behavior of the EoS and therefore might potentially have a sizeable impact on global neutron star properties.  

\begin{table}[]
    \centering
    \begin{tabular}{c|cc}
        \hline
        \hline
         & Min & Max\\
         \hline
          $\rho_{sat}$(fm$^{-3})$ & 0.140 & 0.170 \\
          E$_{sat}(MeV)$ & -17 & -14\\
          K$_{sat}(MeV)$ & 150 & 350\\
          J$_{sym}(MeV)$ & 20 & 60\\
          K$_{sym}(MeV)$ & -300 & 300\\
          b$_\sigma$ & 1.8 & 2.4 \\
          c$_\sigma$ & 2.0 & 4.0 \\
          d$_\sigma$ & 2.8 & 4.2 \\
          b$_\omega$ & 2.0 & 2.4\\
          c$_\omega$ & 2.0 & 3.0\\
          b$_\rho$ & 4.0 & 6.0\\
          d$_\rho$ & -1.0 & 0.0\\
          \hline
          \hline
    \end{tabular}
    \caption{Values used for the generation of the prior distribution of the model parameters $\{\bf X\}$.}
   \label{Tab_generation}
\end{table}

\section{Bayesian Analysis}\label{Sec_Bayes}
In this section we present the details of our Bayesian analysis.

\subsection{Construction of the Prior}

The prior used in this Bayesian analysis consists in a set of GDFM models, each built using the procedure showed in the previous section. The choice of the parameter space to  be explored is a delicate question, because a too restrictive prior will result in biased results for our predictions \cite{Margueron:2017eqc,Char_2023}. To correctly select the couplings parameter space that corresponds to acceptable models from the microphysics point of view, we take advantage of the fact that analytical relations exist \cite{Glendenning:1991es,Dutra:2014qga} between the unknown meson couplings $g_i$ of Eq.(\ref{eq:couplings}) and the so-called empirical nuclear matter parameters (NMPs), that express the physical properties of nuclear matter close to the saturation point that can be explored by laboratory experiments.
Specifically, we will use the definition of the  saturation point of symmetric nuclear matter $\cal{E}(\rho,0)$ (energy $E_{sat}$ and density $\rho_{sat}$), the incompressibility $K_{sat}$, and two parameters characterizing the density behavior of the symmetry energy, namely its value at saturation $J_{sym}$ and its curvature $K_{sym}$. In principle, another parameter that is (at least partially) constrained by nuclear physics is the slope at saturation of the symmetry energy $L_{sym}$. However, we have not directly sampled this parameter since, as we will explicitly show later, the model imposes a strong correlation between the NMPs and it turns out that it is not possible to sample independently $J_{sym}$, $L_{sym}$ and $K_{sym}$. A physical strong correlation between $J_{sym}$ and $L_{sym}$ is found by many authors to arise from the comparison to nuclear data (see for instance \cite{Margueron:2018eob} and references therein), therefore we consider that it is more important to keep the largely unknown  
 $K_{sym}$ as an independent parameter.

The relation between the NMPs and the couplings can be calculated as \cite{Glendenning:1991es,Dutra:2014qga}:

\begin{align}
    E_{sat}=\Bigg[ E_{kin,n}&+E_{kin,p}+\frac{1}{2}m_\sigma^2\phi_0^2 \notag\\
    &+\frac{1}{2}m_\omega^2\omega_0^2+\frac{1}{2}m_\rho^2b_{3,0}^2\Bigg]\Bigg\vert_{\rho=\rho_{sat},\rho_3=0}
    \label{eq.Esat}
\end{align}
\begin{align}
    P_{sat}=0=\Bigg[P_{kin,n}+P_{kin,p}&-\frac{1}{2}m_\sigma^2\phi_0^2+\frac{1}{2}m_\omega^2\omega_0^2 \notag\\ &+\frac{1}{2}m_\rho^2b_{3,0}^2+\rho\Sigma_R\Bigg]\Bigg\vert_{\rho=\rho_{sat},\rho_3=0}
    \label{eq.rhosat}
\end{align}
\begin{align}
    K_{sat}=9\Bigg[\rho\frac{\partial\Sigma_R}{\partial\rho}0+&\frac{2g_\omega\rho^2}{m_\omega^2}\frac{\partial g_\omega}{\partial\rho}+\frac{g_\omega^2\rho}{m_\omega^2}\notag\\
    &\frac{k_F^2}{3E_F}+\frac{\rho M^*}{E_F}\frac{\partial M^*}{\partial\rho}\Bigg]\Bigg\vert_{\rho=\rho_{sat},\rho_3=0}
    \label{eq.Ksat}
\end{align}
\begin{align}
    J_{sym}=\frac{k_F^2}{6E_F}+\frac{g_\rho^2}{8m_\rho^2}\rho\Bigg\vert_{\rho=\rho_{sat},\rho_3=0}
    \label{eq.Jsym}
\end{align}
\begin{align}
K_{sym}=9\rho^2\Bigg\{  
-\frac{\pi^2}{12{E_F^*}^3k_F}&\left(\frac{\pi^2}{k_F} 
+ 2M^*\frac{\partial M^*}{\partial\rho}\right) - \frac{\pi^4}{12E_F^*k_F^4} \notag\\+\frac{g_\rho}{2m_\rho^2}\frac{\partial g_\rho}{\partial\rho}-\Bigg[\frac{\pi^4}{24{E_F^*}^3k_F^2} -
&\frac{k_F\pi^2}{8{E_F^*}^5}\left(\frac{\pi^2}{k_F} +
2M^*\frac{\partial M^*}{\partial\rho}\right)\Bigg] \cdot \notag \\
\Bigg(1 + \frac{2M^*k_F}{\pi^2}\frac{\partial M^*}{\partial\rho}&\Bigg) 
+ \frac{\rho}{4m_\rho^2}\left(\frac{\partial g _\rho}
{\partial\rho}\right)^2-\frac{k_F\pi^2}{12{E_F^*}^3}\cdot\notag\\
\Bigg[\frac{M^*}{k_F^2}\frac{\partial
M^*}{\partial\rho} + \frac{2k_F}{\pi^2}&\left(\frac{\partial
M^*}{\partial\rho}\right)^2 +\frac{2k_F M^*}{\pi^2}\frac{\partial^2
M^*}{\partial\rho^2}\Bigg]\notag\\
&+\frac{g_\rho\rho}{4m_\rho^2}\frac{\partial^2 g_\rho}{\partial\rho^2}\Bigg\}\Bigg\vert_{\rho=\rho_{sat},\rho_3=0}
\label{eq.Ksym}
\end{align}
where $E_F=\sqrt{k_F^2+M^{*2}}$ is the isospin independent nucleon Fermi energy.
 We underline that Eq.\ref{eq.rhosat} actually connects the saturation density of symmetric matter to the couplings by imposing the vanishing of pressure at that specific density.

 Those parameters are sampled using uniform distributions within the large, but physically motivated, intervals given in Table \ref{Tab_generation}.
 The average values and interval extension are chosen such as to fully cover the present uncertainties, while the uniform uncorrelated distributions guarantee that no artificial bias is induced by the largely arbitrary functional form of the couplings\cite{Margueron:2017eqc}. In particular, it was discussed in ref.\cite{Char_2023} 
that a partial or biased exploration of the possible variations of the badly constrained $K_{sym}$ parameter might result in an incomplete description of the lepton fraction distribution at high density, with potential important consequences on the cooling dynamics of neutron stars.
 
 Concerning the seven remaining parameters needed to fully specify each realization of the GDFM model, we again use agnostic uniform distributions with ranges reported in Tab.\ref{Tab_generation}. These intervals are centered on the original values of the GDFM model \cite{Gogelein:2007qa}, and chosen large enough to cover the very largely spread values of the high order NMPs obtained from different relativistic and non-relativistic models in the literature \cite{Margueron:2017eqc}, following the study of ref.
 \cite{Char_2023}.

 The set of models is then filtered so that only the ones that fulfill the following requests are kept:  the fit to the nuclear mass table \cite{Wang_2017} must give a minimum $\chi^2$, which is necessary for the estimation of the surface tension parameter $\sigma$, the variational equations associated to the crust must have a solution within the physical possible values for the crustal composition, 
 and both the uniform neutron matter and the final $\beta$-equilibrated EoS must be thermodynamically stable. Causality is  by construction guaranteed by the relativistic structure of the functional (and a-posteriori verified for each model).
 The $\chi^2$ cited above is defined following\cite{Carreau:2019zdy,DinhThi2023,DinhThi2023a} :
\begin{eqnarray}
	\chi^2_{AME}({\bf X})&=&\frac{1}{N}\sum_{n=1}^N \frac{\left 
	( M_{CLD}^{(n)}({\bf X})-M_{AME}^{(n)}\right )^2}
	{\sigma_{BE}^2} \, ,
        \label{eq_chi2}
\end{eqnarray}
with $N=2408$ the total number of masses considered, $M_{CLD}^{(n)}$ the mass of nucleus $n$ obtained from the best fit of the experimental table within the $\bf X$ parameter set, $M_{AME}^{(n)}$ the experimental value, and $\sigma_{BE}$ chosen 
as 2\% of the corresponding mass, to approximately account for the systematic error of the simplified CLD description. 
 Applying these conditions we are left with a set of roughly $N_{tot}\approx 4\times 10^{5}$, 

 which form what from now on will be referred to as our prior distribution. 

\subsubsection{Prior Re-Weighting}

\begin{figure*}
    \centering
       \includegraphics[width=0.8\linewidth,angle=0]{ 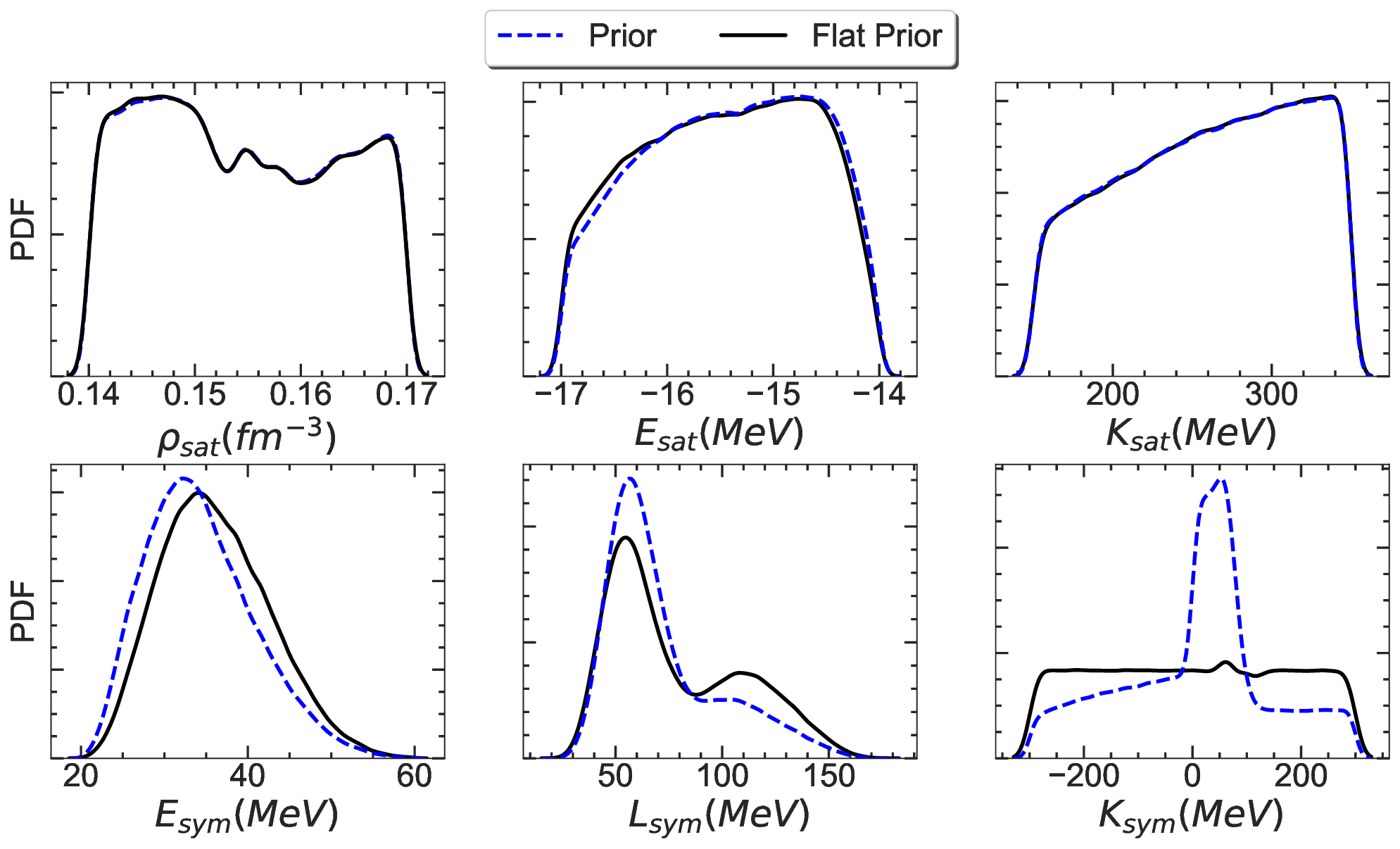}   
    \caption{Marginalized posterior of $\rho_{sat}$ (top left), $E_{sat}$ (top center), $K_{sat}$ (top right), $E_{sym}$ (bottom left), $L_{sym}$ (bottom center) and $K_{sym}$ (bottom right) before (dashed blue) and after (solid black) the re-weighting of the prior}.
    \label{Fig_Ksym_NMP}
\end{figure*}

Even if in the first generation of the model parameters we assume uncorrelated flat distributions, the applied conditions on the behavior of the functional in the crustal region mentioned above obviously result in a slight deformation of the prior NMPs marginalized distributions. In particular, the distribution of $K_{sym}$ results peaked around small and positive values, as it can be seen from panel 6 in Fig.\ref{Fig_Ksym_NMP}.
Though physical correlations among the NMPs are expected from the Lagrangian structure of the functional, we cannot a-priori exclude that some bias might arise from the chosen behavior of the $\rho$-meson coupling and/or the absence of the $\delta$ degree of freedom \cite{Li:2022okx}. 
A fully unbiased prior distribution of $K_{sym}$ might be important to obtain general predictions, since 
in \cite{Reed:2023cap} it was pointed out that the value of the $K_{sym}$ parameter can be crucial in order to reconcile the results of the PREX and CREX experiments \cite{PREX:2021umo,CREX:2022kgg}.  
For this reason, we apply a re-weighting procedure to our prior \cite{PhysRevLett.91.202701}, to specifically ensure the flatness of the distribution of $K_{sym}$. In order to do so, the whole range of $K_{sym}$ from $K_{sym}=K_{sym}^{min}$ to $K_{sym}=K_{sym}^{max}$ is divided in $n$ bins of equal size $\delta K=(K_{sym}^{max}-K_{sym}^{min})/n$, and to each model $\bf X$ is assigned a weight given by
\begin{equation}
    w_{\bf X}= \frac{P_{uniform}}{P_{prior}(i)}
\end{equation}
where $P_{uniform}= 1/n$ and $P_{prior}(i)=N(i)/N_{tot}$ is the number of models in the $i^{th}$ bin of $K_{sym}$, which is the one that contains the model, normalized to the total number of models in the prior. This procedure does not produce important bias on the marginalized distributions of the other parameters, provided that the weight assigned to the models does not vary over orders of magnitude. 

In the present case, the effect of the prior modification on different NMPs marginalized distributions is shown in Fig.\ref{Fig_Ksym_NMP}, as well as in Tab.\ref{Tab_NMP_prior} and Tab\ref{Tab_astro_prior}, where we show the mean and the extremes of the $90\%$ CI for the distributions of the NMPs and of some NS observables respectively.  We can see that the potential bias on the $K_{sym}$ distribution is eliminated by the reweighting procedure, while the other distributions are only very slightly affected. This shows once again that our version of the GDFM functional is sufficiently flexible to describe different possible behaviors of the symmetry energy, without a-priori correlations among the different parameters.  In light of this result, from now on we will always use as a starting point the flat $K_{sym}$ prior, simply calling it "prior".

\begin{table}[]
    \centering
    \begin{tabular}{c|cc}
        \hline
        \hline
         & $\mu$ & $\sigma$\\
         \hline
          $\rho_{sat}$(fm$^{-3})$ & 0.153 & 0.005 \\
          E$_{sat}(MeV)$ & -15.8 & 0.3\\
          K$_{sat}(MeV)$ & 230 & 20\\
          J$_{sym}(MeV)$ & 32.0 & 2.0\\
          K$_{\tau}(MeV)$ & -400 & 100\\
          \hline
          \hline
    \end{tabular}
    \caption{Values of the mean and standard deviation for the NMPs used in the evaluation of $P(Exp|\bf X)$ taken from \cite{Margueron:2017eqc}.}
   \label{Tab_NMP_constraint}
\end{table}

\begin{table*}
    \hskip-2.0cm
    \centering

    \begin{tabular}{cccc|ccc}
    \hline
    \hline 
    &   \multicolumn{3}{c}{Prior} &   \multicolumn{3}{c}{Flat Prior} \\

    & \multicolumn{1}{c}{} & \multicolumn{2}{c}{90\% CI} & \multicolumn{1}{c}{} & \multicolumn{2}{c}{90\% CI} \\
    \cline{3-4} 
    \cline{6-7} 
    & \multicolumn{1}{c}{Mean} & \multicolumn{1}{c}{Min} & \multicolumn{1}{c}{Max} & \multicolumn{1}{c}{Mean} & \multicolumn{1}{c}{Min} & \multicolumn{1}{c}{Max} \\
    \hline
      $\rho_{sat}$(fm$^{-3})$  & 0.154 & 0.141 & 0.168 & 0.154 & 0.141 & 0.168 \\
      $E_{sat}(MeV)$   & -15.4 & -16.7 & -14.2 &-15.5 & -16.8 & -14.3 \\ 
      $K_{sat}(MeV)$   &  260.3 & 164.9 & 341.8 & 260.3 & 165.0 & 341.9 \\ 
      $Q_{sat}(MeV)$   & 386.6 & -964.6 & 1866.7 & 374.7 & -978.1 & 1856.6 \\ 
      $Z_{sat}(MeV)$   & 7476.7 & -6197.4  & 21741.6 & 7409.9 & -6300.1  & 21754.3 \\ 
      $E_{sym}(MeV)$   & 34.3 & 24.9 & 46.2 & 36.1 & 26.3 & 47.7 \\ 
      $L_{sym}(MeV)$   & 71.7 & 41.4 & 127.5 & 78.5 & 40.3 & 136.0 \\ 
      $K_{sym}(MeV)$   & 6.9 & -242.7 & 245.7 & -0.1 & -270.0 & 270.0 \\ 
      $Q_{sym}(MeV)$   & 517.9 & -25.6 & 1195.1 & 609.2 & -21.5 & 1297.3 \\ 
      $Z_{sym}(MeV)$   & -8674.5 & -14767.5 & 334.1 & -7168.7 & -13989.7 & 2641.6 \\
      $K_{\tau}^*(MeV)$   & -423.5 & -631.6 & -250.1 & -471.1 & -662.1 & -284.3 \\ 
      $K_{\tau}(MeV)$   & -523.1 & -1064.7 & -78.5 & -576.2 & -1147.8 & -90.4 \\ 
    \end{tabular}
    \caption{Mean and extremes of the 90\% quantile for the nuclear matter parameters, as well as for the combinations $K_{\tau}^*=K_{sym}-6L_{sym}$ and for $K_{\tau}=K_{sym}-6L_{sym}-Q_{sat}L_{sym}/K_{sat} $. We show the values for the prior and the re-weighted prior with flat $K_{sym}$.}
   \label{Tab_NMP_prior}
\end{table*}

\begin{table*}
    \hskip-2.0cm
    \centering

    \begin{tabular}{cccc|ccc}
    \hline
    \hline 
    &   \multicolumn{3}{c}{Prior} &   \multicolumn{3}{c}{Flat Prior} \\

    & \multicolumn{1}{c}{} & \multicolumn{2}{c}{90\% CI} & \multicolumn{1}{c}{} & \multicolumn{2}{c}{90\% CI} \\
    \cline{3-4} 
    \cline{6-7}  
    & \multicolumn{1}{c}{Mean} & \multicolumn{1}{c}{Min} & \multicolumn{1}{c}{Max} & \multicolumn{1}{c}{Mean} & \multicolumn{1}{c}{Min} & \multicolumn{1}{c}{Max} \\
    \hline
      $R_{1.4M_\odot}(km)$  & 13.7 & 12.6 & 14.9 & 13.8 & 12.6 & 15.1   \\
      $\Lambda_{1.4M_\odot}$   & 1294 & 744 & 2007 & 1336 & 724 & 2156 \\ 
      $\rho^{c}_{1.4M_\odot}$(fm$^{-3})$   & 0.32 & 0.26 & 0.40 & 0.32 & 0.25 & 0.40 \\
      $R_{2.0M_\odot}(km)$   &  13.8 & 12.6 & 14.9 & 13.9 & 12.6 & 15.1 \\ 
      $\Lambda_{2.0M_\odot}$   & 148 & 72 & 241 & 152 & 72 & 256 \\ 
      $\rho^{c}_{2.0M_\odot}$(fm$^{-3})$   & 0.39 & 0.31 & 0.52 & 0.39 & 0.30 & 0.52 \\
      $M_{Max}(M_\odot)$   & 2.57 & 2.32  & 2.85 & 2.58 & 2.33  & 2.87 \\ 
      $\rho^{c}_{M_{Max}}$(fm$^{-3})$   & 0.73 & 0.62 & 0.89 & 0.72 & 0.61 & 0.89 \\ 
    \end{tabular}
    \caption{Mean and extremes of the 90\% quantile for the radius, tidal deformability  and central density of a $1.4M_\odot$ and $2.0M_\odot$ NS, as well as the maximum mass and the central density of the most massive star. We show the values for the prior and the re-weighted prior with flat $K_{sym}$. We underline that the properties of the $1.4M_\odot$ and $2.0M_\odot$ NS are calculated only on those models that reach the respective mass.}
   \label{Tab_astro_prior}
\end{table*}

\subsection{Posterior Calculation}

 The posterior probability distributions of the set 
$\bf X$ of  EoS   parameters is conditioned by 
likelihood models of the different observations and constraints $\mathbf c $ with 
normalizing constant $\mathcal N$ as:
\begin{equation}
        P({\mathbf X}|{\mathbf c})=\mathcal N   \prod_k P(c_k|{\mathbf X}). 
\end{equation}

The corresponding probability distributions for the observables $Y({\bf X})$ 
are obtained by an overall marginalization through the range of values of 
parameters $\bf X$ between ${\bf X}_{min}$ and ${\bf X}_{max}$ according to
\begin{equation}
        P(Y|{\mathbf c})=\prod_{k=1}^N\int_{X_k^{min}}^{X_k^{max}}dX_k \, 
	P({\mathbf X}|{\mathbf c}) \delta\left (Y-Y({\mathbf X})\right ).
\end{equation}

As already mentioned, in our study we consider the effect of three main categories of constraints.  \\ The first category consists in the theoretical constraints on the properties of neutron matter coming from ab-initio calculations based on $\chi$-EFT, the second category consists in the experimental constraints on the properties of nuclear matter close to saturation, and the last one includes different astrophysical constraints on the observables of NS. In the following sub-sections we  detail the likelihood models $P(c_k|{\mathbf X})$ for the three categories.

\subsubsection{Theoretical nuclear constraints}\label{section:theo}

In our study we consider the ab-initio calculation of the energy per baryon of neutron matter, taken from the compilation presented in the left part of Fig.1 in ref.\cite{Huth2021}. 
This compilation comprises different many-body calculations using interactions from  $\chi$-EFT, namely many-body perturbation theory \cite{Hebeler_2013,Tews_2013,Drischler_2019,Drischler:2021kxf}, and auxiliary field diffusion Monte-Carlo \cite{Lynn_2016}.  The ensemble of these calculations leads to a band in the energy-density plane, that comprises EFT truncation errors and different regulators, uncertainties in the low-energy couplings that enter three-nucleon forces, as well as systematics due to the different many-body methods.  
 As such, the band represents the region of compatibility of a given nuclear model with our theoretical knowledge of the low-energy nucleon-nucleon interaction. However, the expression of a likelihood model to quantify the meaning of the band in a probabilistic way is not unique \cite{Drischler_2020}, therefore 
this constraint will be implemented in two different ways, detailed below, which will then be compared.

\begin{itemize}
    \item {\it "Heavyside"}: In the first case, we assume a simple pass-band filter, where we consider the band given by \cite{Huth2021} as a 90\% confidence interval, as previously done in similar studies\cite{Carreau:2019zdy,universe7100373}. In this case we will divide the constraint into N baryon density slices and the weight will be given by 

    \begin{eqnarray} \label{eq:filter}
        &&P(\chi\text{EFT},HS|{\mathbf X})\\ \nonumber
        && \;\;\;\propto \prod_{i=1}^{N} \begin{cases} 1 & \mbox{if} \;
        x_{min}^i-\delta^i < x_i({\bf X}) < x_{max}^i+\delta^i \\ 0 & 
        \mbox
        {otherwise} \end{cases},
    \end{eqnarray}
    where  $x_i(\bf X)=\mathcal{E}(\rho_i,\rho_i)/\rho_i$ is the energy per baryon in pure neutron matter of the model $\bf X$ at the $i^{th}$ point in density, $x_{min}^i$ and $x_{max}^i$ are the lower and upper bound of the theoretical band at the same density point and $\delta^i=0.05(x_{max}^i-x_{min}^i)$.

    \item {\it "Gaussian"}: In the second case we consider a weaker way to impose the constraint. Once again we slice the density domain where the constraint is applied into N equal intervals, and we assume that all the models that fall within the range of the band in all density intervals will have equal weight. However, unlike the previous case, the models that do not fall inside the band in one or more density intervals are not discarded. The energy per baryon at the different density points are rather treated as independent Gaussian variables. The likelihood becomes:   
    \begin{eqnarray} \label{eq:gaussian}
        &&P(\chi\text{EFT},Gauss|{\mathbf X})\\ \nonumber
        && \;\;\;\propto \prod_{i =1}^{N} \begin{cases} P_U^i(x_i) & \mbox{if } x_{min}^i < x_i({\bf X}) < x_{max}^i \\ P_G^i(x_i) & \mbox{otherwise} \end{cases} ,
    \end{eqnarray}
    where $x_i$, $x_{min}^i$ and $x_{max}^i$ are defined as in the previous case and where
    \begin{equation}
        P_U^i(x)=\frac{0.682}{2\sigma_i},
    \end{equation}
    while
    \begin{equation}
        P_G^i(x)=\frac{1}{\sigma_i\sqrt{2\pi}}e^{-\frac{1}{2}(\frac{x'-\mu_i}{\sigma_i})^2},
    \end{equation}
    with $\mu_i=(x^i_{max}+x^i_{min})/2$ and $\sigma_i=(x^i_{max}-x^i_{min})/2$.

\end{itemize}

\subsubsection{Experimental nuclear constraints}

In this category we consider 
indirect constraints on the NMPs coming from the interpretation of nuclear experimental data through different versions of the density functional theory, see \cite{sorensen2023dense} for a recent review. We consider a gaussian likelihood model on the five NMPs that are better constrained by nuclear experiments, namely $\rho_{sat}$, E$_{sat}$, K$_{sat}$ and J$_{sym}$ \cite{Margueron:2017eqc}, as well as on the combination $K_{\tau}=K_{sym}-6L_{sym}-Q_{sat}L_{sym}/K_{sat}$, that was shown in refs.\cite{Piekarewicz_Centelles,Khan2013} to be well correlated to experimental data on the isoscalar giant monopole resonance. These variables are considered as independent, and the likelihood is given by: 
\begin{eqnarray}
    P(Exp|{\mathbf X})
      =  \prod_{i=1}^{5}  \frac{1}{\sigma_i\sqrt{2\pi}}e^{-\frac{1}{2}(\frac{x'-\mu_i}{\sigma_i})^2}.
\end{eqnarray}
Here, the five values of $i$ refer to the four NMPs listed above, namely 
$x_1=\rho_{sat}({\bf X})$, $x_2=E_{sat}({\bf X})$, $x_3=K_{sat}({\bf X})$, $x_4=J_{sym}({\bf X})$, $x_5=K_{\tau}({\bf X})$ and  the values of $\mu_i$ and $\sigma_i$ for each NMP are taken from the compilation of experimental data in ref. \cite{Margueron:2017eqc} and showed in Tab.\ref{Tab_NMP_constraint}.

\begin{figure*}
    \centering
       \includegraphics[width=0.8\linewidth,angle=0]{ 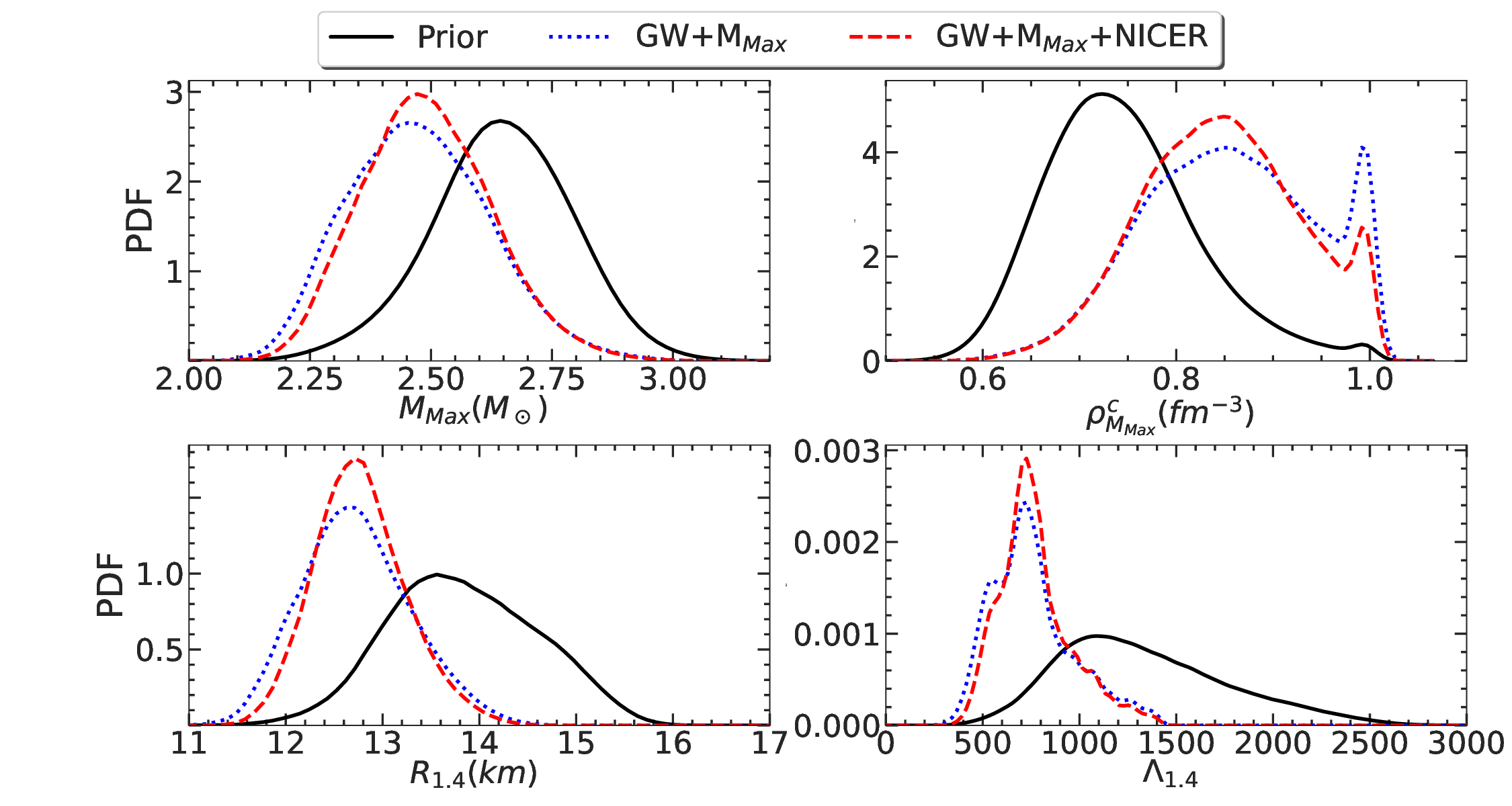}   
    \caption{Marginalized posterior of the maximum mass (top left), the central density of the most massive star (top right) and of the radius (bottom left) and tidal deformability (bottom right) of a $1.4M_\odot$ NS for the prior distribution (solid black), the posterior with only the maximum mass and GW170817 constraints (dotted blue) and the full astrophysical constraint (dashed red)}.
    \label{Fig_astro_astro}
\end{figure*}

\subsubsection{Astrophysical Constraints}

The likelihood associated to the astrophysical constraints is given by the product of three terms: a constraint on the maximum mass of neutron stars coming from the precise measurement of very massive pulsars \cite{Antoniadis:2013pzd,Arzoumanian:2017puf,Fonseca:2021wxt}, the constraint on the tidal deformability given by the GW170817 event measured by the Ligo and Virgo collaboration \cite{TheLIGOScientific:2017qsa,Abbott_2017_2,Abbott_2019}, and two constraints on the NS radius coming from the NICER observations \cite{Riley:2019yda,Miller:2019cac,Riley:2021pdl,Miller:2021qha}. The  posterior is thus conditioned by a total likelihood expressed as:
\begin{equation}
    P(Astro|\textbf{X})=P(M_{max}|\textbf{X})\cdot P(LVC|\textbf{X})\cdot P(NICER|\textbf{X}).
\end{equation}

For the maximum mass constraint, we use the precise Shapiro delay radio timing measurement of the pulsar J0348+0432 \cite{Antoniadis:2013pzd}, with an estimated mass  $(2.01\pm 0.04) M_\odot$. The likelihood  is calculated as the cumulative distribution function of a Gaussian distribution centered at $2.01$ and with a standard deviation of $0.04$. The wight is thus defined as 
\begin{equation}
        P(M_{max}|\textbf{X})= \frac{1}{0.04\sqrt{2\pi}}\int^{M_{max}({\bf X})/M_\odot}_{0}e^{-\frac{1}{2}(\frac{x-2.01}{0.04})^2}dx ,
\end{equation}
where  $M_{max}({\bf X})$ is the maximum NS mass at equilibrium, determined from the solution of the TOV equations.
Concerning the constraint coming from the observation of GW170817 \cite{TheLIGOScientific:2017qsa,Abbott_2017_2,Abbott_2019}, we fix the value of the chirp mass to be $\mathcal{M}=1.186M_\odot$, neglecting the (very small) uncertainty on this quantity. We then vary the quantity $q=m_2/m_1$, considering N steps in the range $[0.57,1.00]$ and assign a weight to each model using the two-dimensional posterior distribution  $P(\tilde{\Lambda}(q),q)$ reported in  ref.\cite{TheLIGOScientific:2017qsa} using the PhenomPNRT waveform:
\begin{equation}
        P(LVC|\textbf{X})= \frac{1}{n_p} {\sum_{i=1}^{n_p} P(\tilde{\Lambda}(q_i),q_i)}
\end{equation}
where $n_p\leq N$ is the number of steps in $q$ which are allowed by the maximum mass of the specific model.\\
Finally, the weight coming from the constraint on the radius given by the NICER observations is given by the product of the weight coming from the observation of PSR J0030+0451 \cite{Riley:2019yda,Miller:2019cac} and the one of PSR J0740+6620 \cite{Riley:2021pdl,Miller:2021qha}}, so that the total likelihood is given by
\begin{equation}
     P(NICER|\textbf{X})=P_{J0030}(R,M)\cdot P_{J0740}(R,M).
\end{equation}
 Here we use the data from \cite{Riley:2019yda,Riley:2021pdl}.
For the calculation of the two terms, we use the two-dimensional posterior distribution for masses and radii. We consider N steps in the mass $M^i$ between the lower and upper bound of the region given by the constraint and sum all the contribution, obtaining a global likelihood modeled as:
\begin{equation}
        P_{j}(\textbf{X})=\frac{1}{n_p} {\sum_{i=1}^{n_p} P_j^i(M^i,R(M^i))} \, ,
\end{equation}
where $j$ refers to the considered pulsar, and  $n_p\leq N$ is the number of steps in mass which are allowed by the maximum mass of the specific model.

\section{Results and Discussion}\label{Sec_res}
In this section we show the results obtained in our study.\\

\begin{table*}[]
    \hskip-2.0cm
    \centering

    \begin{tabular}{cccc|ccc|ccc}
    \hline
    \hline 
    &   \multicolumn{3}{c}{Prior} &   \multicolumn{3}{c}{GW+$M_{Max}$} & \multicolumn{3}{c}{GW+$M_{Max}$+NICER}\\

    & \multicolumn{1}{c}{} & \multicolumn{2}{c}{90\% CI} & \multicolumn{1}{c}{} & \multicolumn{2}{c}{90\% CI} & \multicolumn{1}{c}{} & \multicolumn{2}{c}{90\% CI}\\
    \cline{3-4} 
    \cline{6-7} 
    \cline{9-10} 
    & \multicolumn{1}{c}{Mean} & \multicolumn{1}{c}{Min} & \multicolumn{1}{c}{Max} & \multicolumn{1}{c}{Mean} & \multicolumn{1}{c}{Min} & \multicolumn{1}{c}{Max} & \multicolumn{1}{c}{Mean} & \multicolumn{1}{c}{Min} & \multicolumn{1}{c}{Max}\\
    \hline
      $\rho_{sat}$(fm$^{-3})$   & 0.154 & 0.141 & 0.168 & 0.160 & 0.145 & 0.169 & 0.160 & 0.145 & 0.169  \\
      $E_{sat}(MeV)$    &-15.5 & -16.8 & -14.3 & -15.5 & -16.8 & -14.3 & -15.5 & -16.8 & -14.3 \\ 
      $K_{sat}(MeV)$    & 260.3 & 165.0 & 341.9 & 233.8 & 158.1 & 328.1 & 233.8 & 157.9 & 328.2 \\ 
      $Q_{sat}(MeV)$    & 374.7 & -978.1 & 1856.6 & -387.5 & -1349.8 & 916.8 & -349.4 & -1302.1 & 892.4\\ 
      $Z_{sat}(MeV)$    & 7409.9 & -6300.1  & 21754.3 & 3650.3 & -8646.0  & 18167.3 & 4200.7 & -8408.2  & 18443.1\\ 
      $E_{sym}(MeV)$    & 36.1 & 26.3 & 47.7 & 36.4 & 26.5 & 47.3 & 36.3 & 26.6 & 47.1 \\ 
      $L_{sym}(MeV)$    & 78.5 & 40.3 & 136.0 & 56.0 & 41.8 & 69.4 & 55.6 & 41.8 & 68.5 \\ 
      $K_{sym}(MeV)$    & -0.1 & -270.0 & 270.0 & -148.1 & -288.4 & 51.9 & -151.7 & -289.0 & 48.8 \\ 
      $Q_{sym}(MeV)$    & 609.2 & -21.5 & 1297.3 & 728.8 & 50.4 & 1380.3 & 723.7 & 44.5 & 1373.0 \\ 
      $Z_{sym}(MeV)$    & -7168.7 & -13989.7 & 2641.6 & -4464.6 & -12062.2 & 4175.2 & -4279.9 & -12023.5 & 4574.8 \\
      $K_{\tau}^*(MeV)$    & -471.1 & -662.1 & -284.3 & -484.3 & -693.4 & -274.4 & -485.0 & -690.0 & -277.4 \\ 
      $K_{\tau}(MeV)$    & -576.2 & -1147.8 & -90.4 & -379.2 & -787.1 & -13.3 & -389.3 & -776.5 & -32.1 \\ 
    \end{tabular}
    \caption{Mean and extremes of the 90\% quantile for the nuclear matter parameters, as well as for $K_{\tau}^*=K_{sym}-6L_{sym}$ and for $K_{\tau}=K_{sym}-6L_{sym}-Q_{sat}L_{sym}/K_{sat} $. We show the values for the prior, the posterior obtained taking into account the constraint on the mass coming from the observation of J0348+0432 and the constraint coming from GW170817 and the posterior obtained including also the constraint coming from NICER.}
   \label{Tab_NMP_astro}
\end{table*}

\subsection{Astrophysical Constraints} \label{sec:astro_posterior}

 \begin{figure*}
    \centering
       \includegraphics[width=0.8\linewidth,angle=0]{ 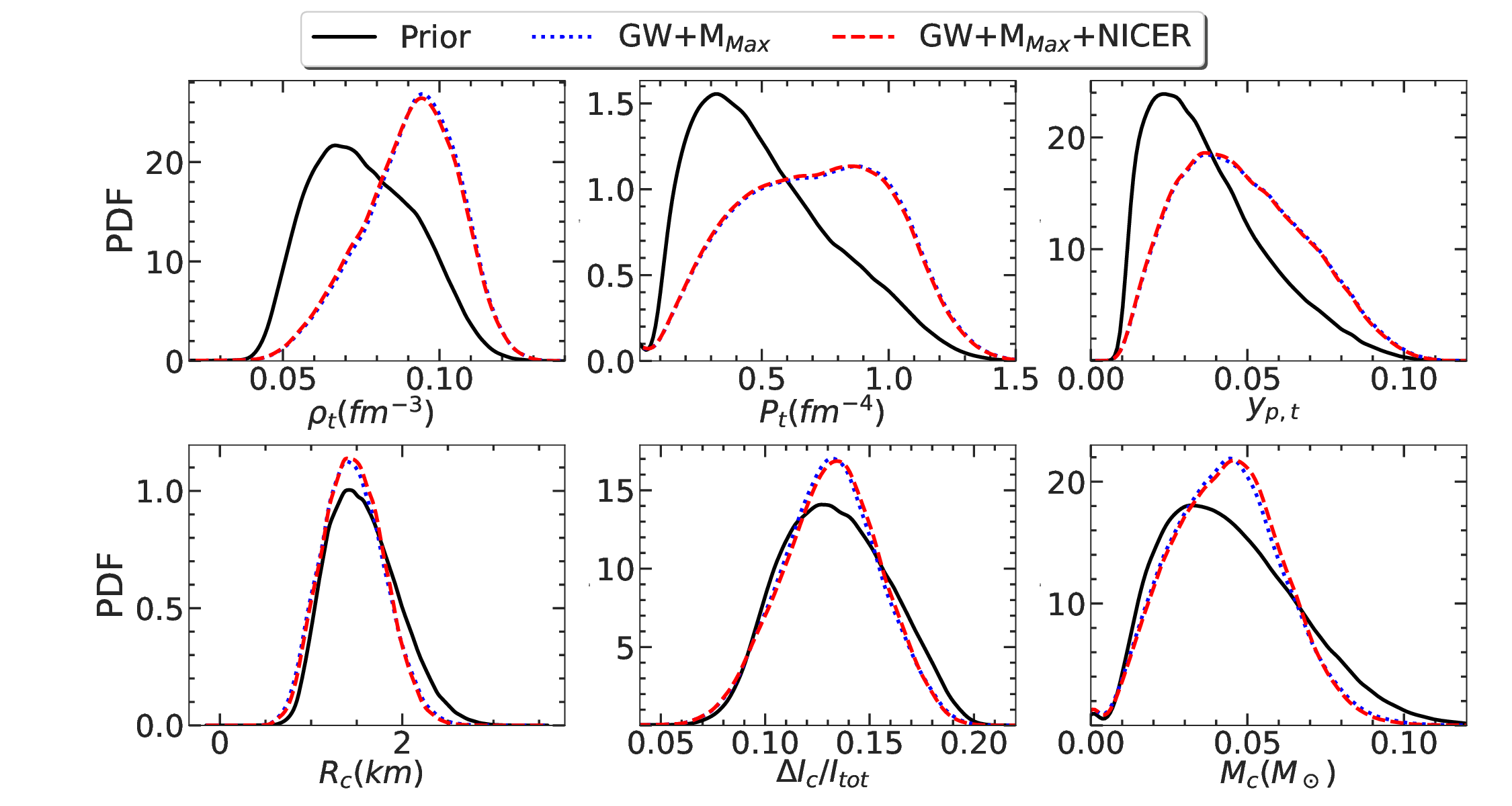}   
    \caption{Marginalized posterior of the crust-core transition density (top left), pressure (top center) and proton fraction (top right), as well as of the crustal radius (bottom left), moment of inertia (bottom center) and mass (bottom right) of a $1.4M_\odot$ NS for the prior distribution (solid black), the posterior with only the maximum mass and GW170817 constraints (dotted blue) and the full astrophysical constraint (dashed red) }.
    \label{Fig_crust_astro}
\end{figure*}

We start our study by analyzing the effect of the constraints coming from the astrophysical observations which were listed in the previous section. We want to study in particular the effect of the constraint coming from the NICER observations of the radius, that were not analyzed in previous works using similar meta-modelling techniques \cite{universe7100373,Mondal:2022cva,Char_2023}. For this reason we compare the two posterior distributions: one in which we only apply the constraint on the maximum mass and the one coming from GW170817 and one in which we apply all the astrophysical constraints. In Fig.\ref{Fig_astro_astro} we show the effect of the astrophysical constraints on some NS observables. It is immediately clear that the constraints have a very strong effect on the observables. In particular, the maximum mass distribution is shifted to lower values, while the central density of the most massive star is shifted to higher values. This result is in qualitative agreement with previous Bayesian analyses with RMF priors \cite{Malik_2022,Char_2023} and can be understood from the fact that the GW170817 measurement favors relatively soft EoSs, corresponding to lower values for the maximum mass and higher values for the central density. It is also apparent from the prior distribution of the maximum mass that the maximum mass constraint, that is extremely important in the non-relativistic version of the nucleonic meta-model\cite{universe7100373,Fantina_2021}, plays virtually no role in our results. This can be understood from the fact that any non-relativistic models become non-causal at relatively low  NS mass, and the implementation of the causality constraint by simple rejection of the non-causal models creates a bias in the maximum mass distribution. In view of this result, we can anticipate that within this relativistic treatment the possible appearence of hyperons in the NS core will not be hindered as it is the case in the non-relativistic models \cite{Oertel:2016bki}, thus potentially eliminating the so called "hyperon puzzle"\cite{BURGIO2021103879}. The inclusion of hyperons in the treatment is relatively straightforward, and it is left for further developements of the GDFM meta-model.

Coming back to Figure \ref{Fig_astro_astro}, we also notice the appearance of a second peak at higher densities in the posterior central density distribution. 
Though it is not clear why the astrophysical constraints should lead to a bimodal distribution of the central density, it turns out that 

these models present values of the NMPs far outside the ranges showed in Table \ref{Tab_NMP_constraint}, with very high saturation densities, low saturation energies and low $K_{sat}$. For this reason, it will be shown later that these models will be discarded by the nuclear experimental constraints.

Concernig the properties of the canonical  $1.4M_\odot$ NS displayed in the lower part of the figure, with no surprise the distribution of the tidal deformability, that is very spread and structureless in the prior, becomes peaked following the constraint imposed by the LVC $(\tilde\Lambda,q)$ measurement of GW170817. The well known correlation between $\Lambda$ and $R$ 
leads to a relatively precise prediction for the radius $R_{1.4}=(12.8\pm 0.5)$ km. For all variables, the addition of the NICER constraints does not significantly modify the posterior distributions. This is another consequence of the good correlation between $\Lambda$ and $R$. Due to this correlation, the relatively precise measurement of the tidal polarizability from GW170817 turns out to constrain the NS radii in a stronger way than the radius measurement from NICER, that is more direct but affected by important systematics. Still, the perfect agreement between the "$GW+M_{max}$" and "$GW+M_{max}+NICER$" posterior shows that these very different measurements coming from completely different probes and concerning different objects are perfectly compatible with each other, as well as with a nucleonic composition of the NS core, in agreement with previous analyses \cite{universe7100373, Fantina_2021,Malik_2022}. In particular, our calculations do not support the need of an EoS stiffening at high density due to the PSR J0740+6620 radius measurement, that was seen in some previous studies employing a more limited set of nucleonic models\cite{Somasundaram:2021ljr}.

\begin{figure*}
  \begin{minipage}[b]{0.49\linewidth}
  \centering
  \includegraphics[width=1.15\linewidth,angle=0]{ 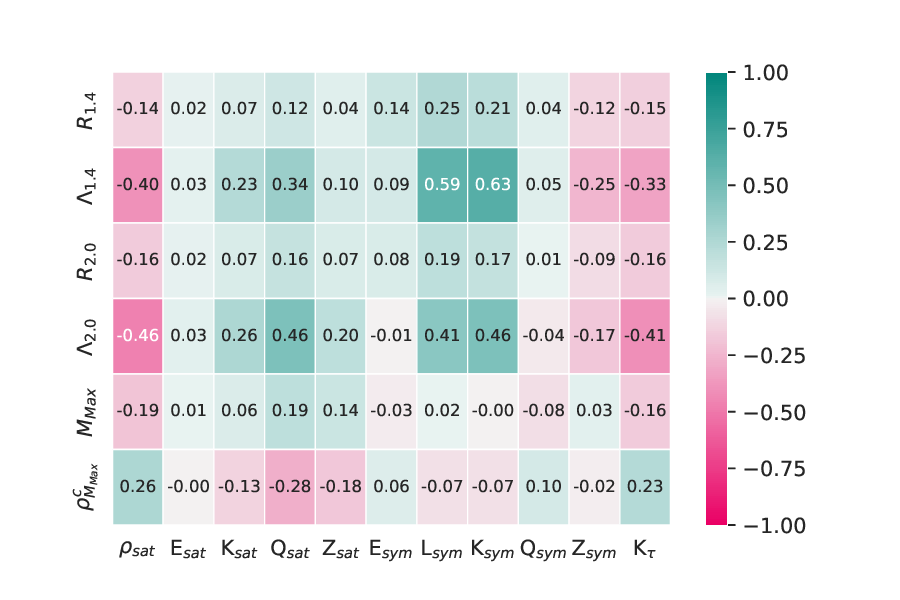}  
  \end{minipage}
  \hfill
  \begin{minipage}[b]{0.49\linewidth}
  \centering
  \includegraphics[width=1.15\linewidth,angle=0,right]{ 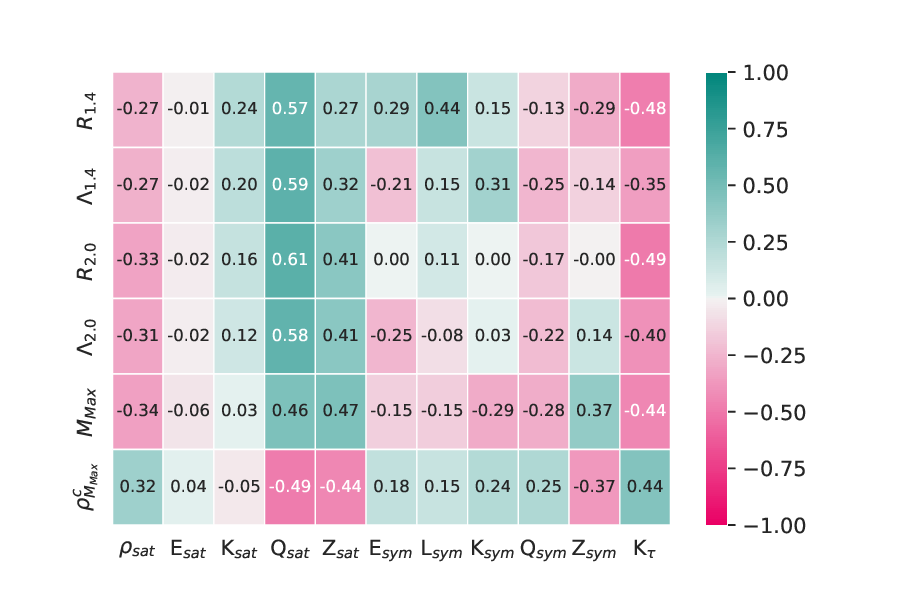}  \end{minipage}
\caption{Linear correlation between the NMPs and the astrophysical observables of NS for the prior distribution (left) and the full astrophysical posterior (right). }
    \label{Fig_corr_astro_astro}
\end{figure*}

\begin{figure*}
  \begin{minipage}[b]{0.49\linewidth}
  \centering
  \includegraphics[width=1.15\linewidth,angle=0]{ 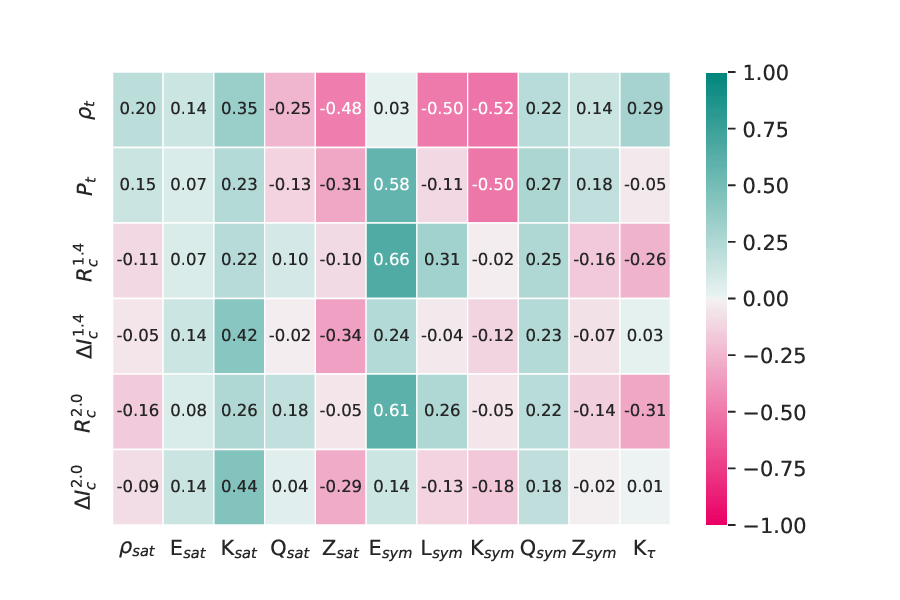}  
  \end{minipage}
  \hfill
  \begin{minipage}[b]{0.49\linewidth}
  \centering
  \includegraphics[width=1.15\linewidth,angle=0,right]{ 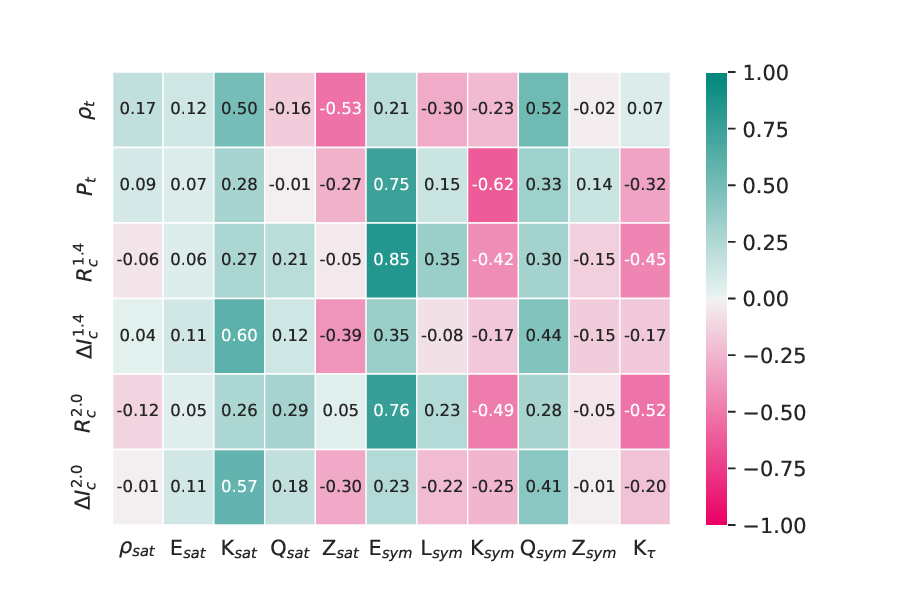}  \end{minipage}
\caption{Linear correlation between the NMPs and the crustal properties of NS for the prior distribution (left) and the full astrophysical posterior (right).}
    \label{Fig_corr_crust_astro}
\end{figure*}

In Fig.\ref{Fig_crust_astro} we show the effect of the  constraints from astrophysical observations on the properties of the NS crust.  As expected, the effect on the crustal mass, radius and moment of inertia is much smaller with respect to the one observed on the astrophysical observables. 
However, a sizeable effect is seen on the properties of the crust-core transition point shown in the upper part of the figure, as previously observed in the non-relativistic version of the meta-model\cite{DinhThi2023,DinhThi2023a}. 
We see that the distribution of the crust-core transition density gets shifted to higher values, while the ones of transition pressure and proton fraction become broader. However, the astrophysical observations concern global properties of the star and do not allow to specifically pin down the behavior of the energy density and pressure at the low densities corresponding to the crust. Because of that, these results are 
mostly related to the correlation with the internal parameters of the RMF model, and in particular are not fully compatible with the ones of refs.\cite{DinhThi2023,DinhThi2023a}. However, we will see that the compatibility will be recovered when we will consider the full posterior including constraints specifically acting on the low density part of the EoS.
Similar to what was observed in Fig. \ref{Fig_astro_astro} above, the most important constraint comes from the GW measurement, while the NICER observations do not bring important additional information.
\\
\begin{figure*}
  \centering
  \hskip-2.0cm
  \includegraphics[width=1.1\linewidth,angle=0]{ 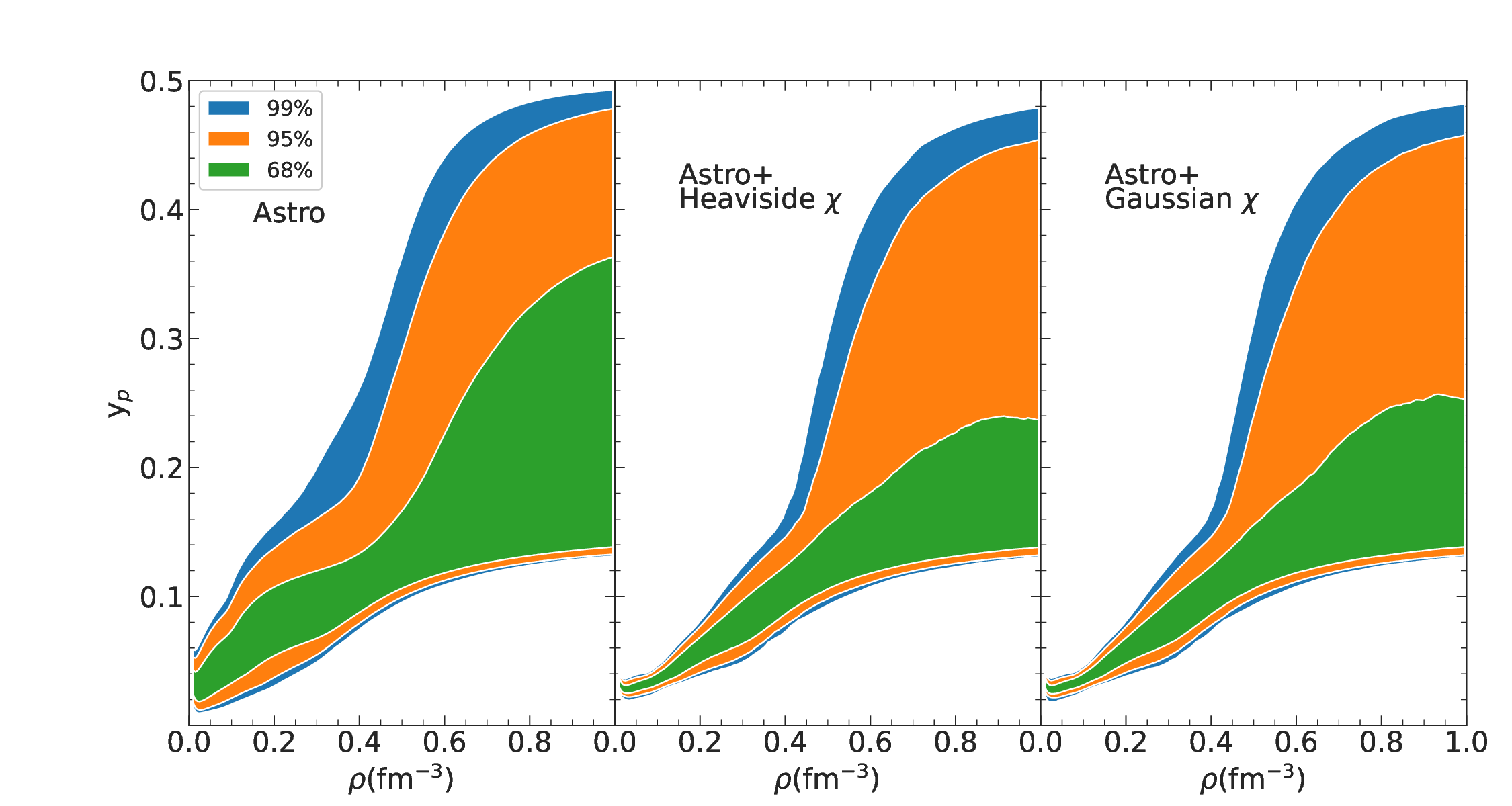}  
\caption{99\%, 95\% and 68\% quantiles for the distribution of the proton fraction as a function of the baryonic density for the astrophysical posterior (left), the posterior with chiral constraint implemented using the likelihood model of Eq.(\ref{eq:filter}) (center) and the  one using  the likelihood model of Eq.(\ref{eq:gaussian})  (right)}
    \label{Fig_yp_filt}
\end{figure*}

We finish the analysis of the astrophysical constraints by studying their effect on the nuclear matter parameters, reported  in Tab.\ref{Tab_NMP_astro}.  We notice that the constraints affect mostly the higher order parameters as expected. Also the distributions of $L_{sym}$ and $K_{sym}$ are affected, shifting the distribution to lower values in the first case and to negative values in the second. The effect on the NMPs and the one on the NS observables can be linked by 
looking at the Pearson correlation coefficient of the NMPs with  the astrophysical observables, displayed in Fig.\ref{Fig_corr_astro_astro}, as well as with the crustal properties,   shown in Fig.\ref{Fig_corr_crust_astro}. 
We can see that non-negligible correlations appear already at the level of the prior, even if it is clear that no NMP can be univocally associated to a NS properties, and both the isoscalar and the symmetry sector contribute  to the different observables. The relatively strong prior correlation between the saturation density $\rho_{sat}$ and the global properties of NS is related to the internal correlations of the RMF models and will be better explained in subsection \ref{sec:models}.

In any case, when the astrophysical constraints are applied, the correlation between the observables and some NMPs drastically increase. In particular all the observables displayed in Fig. \ref{Fig_corr_astro_astro} appear to be strongly correlated to $Q_{sat}$ and $Z_{sat}$, which explains why the distributions of these two parameters are strongly affected by the constraints. On the other hand, the correlations between the NMPs and the crustal properties appear to be less modified when the constraints are applied. As expected, these quantities appear to be more correlated to the lower order parameters, which are less influenced by the constraints. This explains the fact that also the crustal properties are less affected.

\subsection{Effect of the Chiral constraint}

\begin{figure*}
  \centering
  \hskip-2.0cm
  \includegraphics[width=1.1\linewidth,angle=0]{ 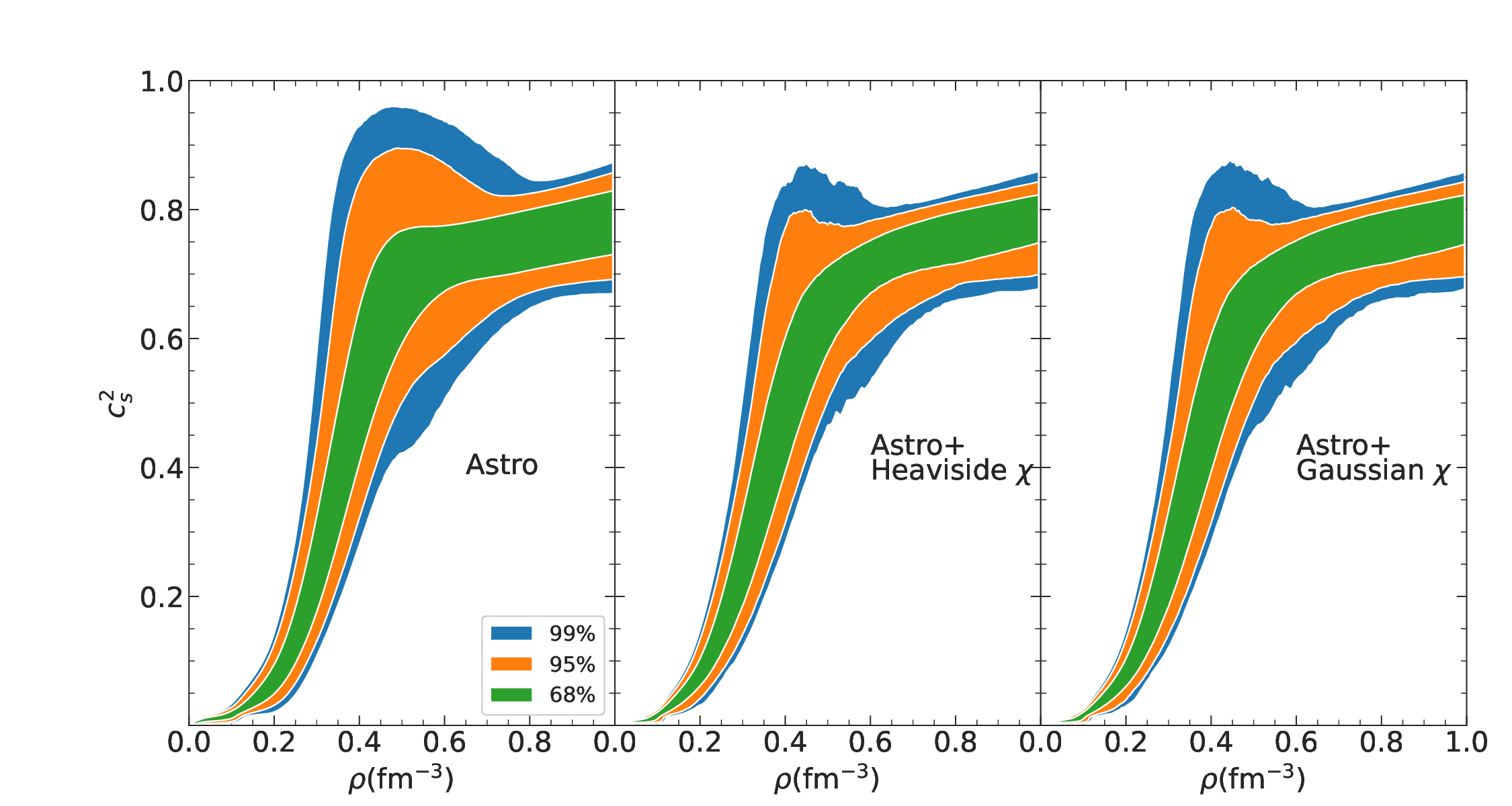}  
\caption{99\%, 95\% and 68\% quantiles for the distribution of the speed of sound as a function of the baryonic density for the astrophysical posterior (left), the posterior with chiral constraint implemented  using the likelihood model of Eq.(\ref{eq:filter}) (center) and the  one using  the likelihood model of Eq.(\ref{eq:gaussian})  (right)}
    \label{Fig_cs_filt}
\end{figure*}

\begin{figure*}
    \centering
       \includegraphics[width=0.8\linewidth,angle=0]{ 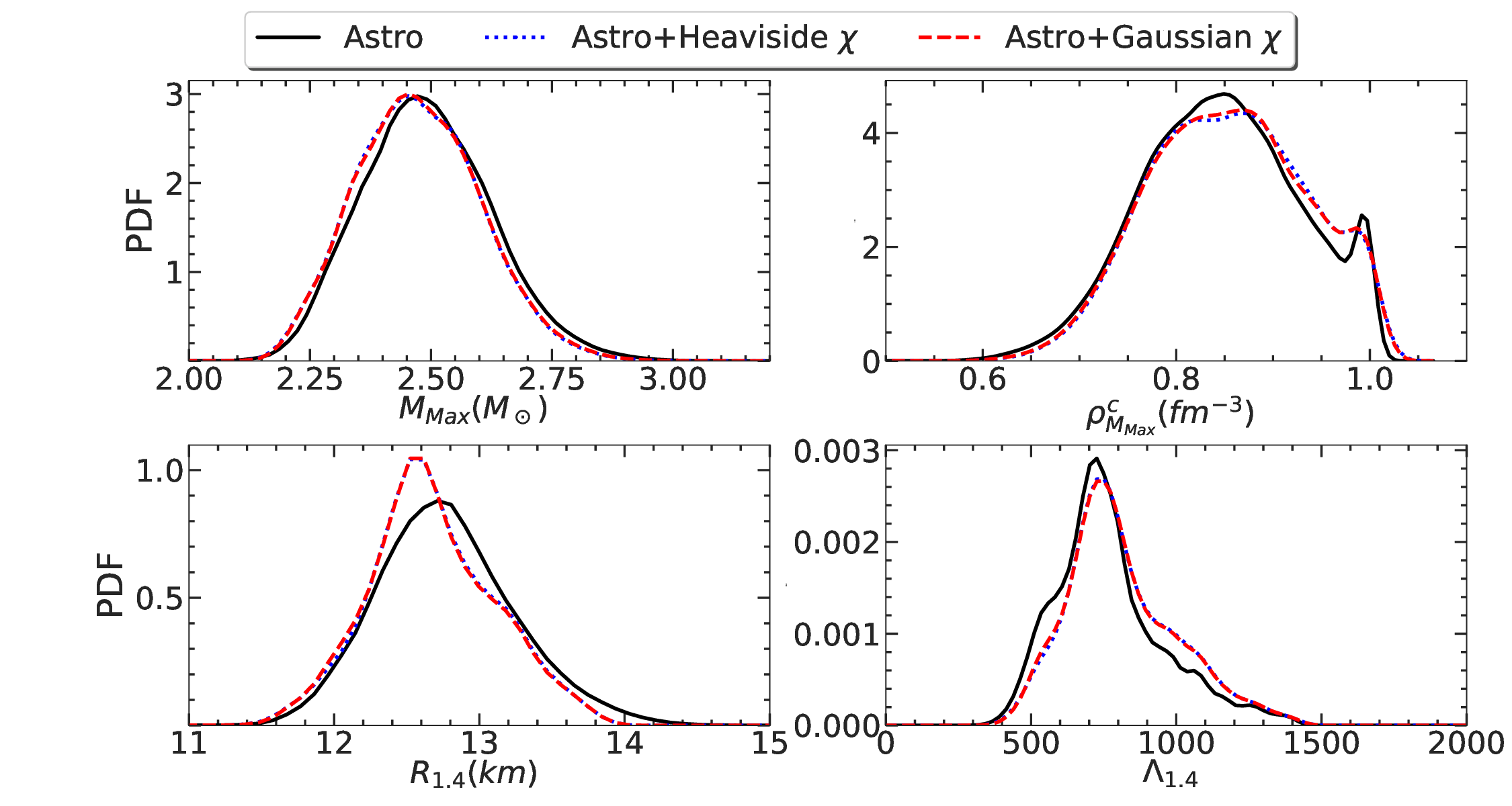}   
    \caption{PDF of the maximum mass (top left), the central density of the most massive star (top right) and of the radius (bottom left) and tidal deformability (bottom right) of a $1.4M_\odot$ NS for the prior distribution (solid black),  the posterior with chiral constraint implemented  using the likelihood model of Eq.(\ref{eq:filter}) (center) and the  one using  the likelihood model of Eq.(\ref{eq:gaussian})  (right)}.
    \label{Fig_astro_chiral_1}
\end{figure*}

After examining the effect of the astrophysical constraints, we continue our analysis by studying the effect of the constraints coming from ab-initio calculations of pure neutron matter, using a large compilation of results from different groups and approaches \cite{Huth2021} that covers the present theoretical uncertainty. This analysis will be divided in two parts: the study of the difference between the two ways ({\it "Heavyside"} and {\it "Gaussian"}) of implementing the constraint introduced in Section \ref{section:theo}, and the study of the influence of the density range in which the constraint is applied.\\
We start by comparing the two ways of implementing the conditional probability. In Fig.\ref{Fig_yp_filt} and Fig.\ref{Fig_cs_filt} we show $99\%$, $95\%$ and $68\%$ quantiles for respectively the proton fraction and the speed of sound squared of beta-equilibrated matter as a function of density. The astrophysical posterior obtained in Section \ref{sec:astro_posterior} (left panels), that here acts as a prior distribution for the nuclear theory constraint, is compared with a posterior distribution additionally conditioned by  $P(\chi\text{EFT},HS|{\mathbf X})$ defined in Eq.(\ref{eq:filter}) (central panels), as well as with 
a posterior conditioned by $P(\chi\text{EFT},Gauss|{\mathbf X})$ defined in Eq.(\ref{eq:gaussian}) (right panels). In all cases, the constraint is applied in the density range  $(0.02-0.2)$ fm$^{-3}$, meaning that we consider in Eqs.(\ref{eq:filter}),(\ref{eq:gaussian}) $\rho_1=0.02$ fm$^{-3}$, $\rho_N=0.2$ fm$^{-3}$.

 For both observables,  it is clear that the two methods lead to  very close  results at all densities, thus indicating 
 the independence on the likelihood model. This  is further confirmed in a more quantitative way in Fig.\ref{Fig_astro_chiral_1}, where we show the marginalized distributions of some selected astrophysical quantities, namely  the maximum mass, the central density of the most massive star and the radius and tidal deformability of a $1.4M_\odot$ NS. Also in this case we can see that the difference between the two methods is negligible.
 In light of this results, from now on, the chiral constraint will always be implemented using the gaussian formulation. 
 
 Concerning the constraining effect of ab-initio nuclear theory, we can see that it is extremely important at low density as expected, by narrowing in a very important way the matter composition in the density region where the theory can be applied. Moreover, it is clear that the constraint is also effective at higher densities than the ones in which it is applied. This is at variance with fully agnostic EoS models such as piecewise polytropes, sound speed models, or gaussian processes \cite{Read:2008iy,Lindblom:2010bb,Lindblom:2012zi,Lindblom:2013kra,Landry:2018prl,Essick:2019ldf,Landry:2020vaw,Legred_2022}, 
 and originates from the unique Lagrangian structure with nucleonic degrees of freedom that is the main hypothesis underlying our predictions. Because of that, once confronted to future observations, our predictions can be used as a null hypothesis to pin down the possible emergence of exotic degrees of freedom in dense matter \cite{Iacovelli_2023}. 
 
 The effect of the filter on the sound speed squared on Fig.\ref{Fig_cs_filt} is less pronounced than the one on the proton fraction  Fig.\ref{Fig_yp_filt}. This is due to the fact that the sound speed is linked to a second order derivative of the energy density, that is very poorly constrained when the energy density is represented by band.  
 Bands representing the uncertainty of present ab-initio calculations of the pressure of neutron matter have also been published \cite{Hebeler_2013,Tews_2013,Lynn_2016,Drischler_2019,Drischler:2021kxf}. However, the propagation of the systematics on the energy density (that is the quantity directly calculated in many body approaches) to the pressure is a subtle issue\cite{Drischler_2020,Wesolowski21}, and when comparing different many-body methods it is not clear if the band comprising the curves derivatives can be interpreted as the uncertainty on the derivative of the energy density band. Progress on the issue of uncertainty propagation in $\chi$-EFT approaches is ongoing in the ab-initio community, and will be of foremost importance for a better understanding of the NS observables. 
 
 Finally, the effect of the nuclear theory constraint on the global astrophysical quantities in Fig.\ref{Fig_astro_chiral_1} is seen to be very small, in agreement with previous results in the literature \cite{universe7100373,Char_2023}. This is due to the fact that the quantities represented in Fig.\ref{Fig_astro_chiral_1} depend on the EoS through the solution of the hydrostatic structure equations in general relativity. As such, they need strong constraints on the pressure over a very wide range of baryonic densities. The peak seen in the central density distribution in Fig. \ref{Fig_astro_chiral_1}, as already seen in Fig.\ref{Fig_astro_astro}, correspond to outliers that will be discarded by the experimental constraints, as it will be discussed in the next section.

\begin{figure*}
  \centering
  \hskip-2.0cm
  \includegraphics[width=1.1\linewidth,angle=0]{ 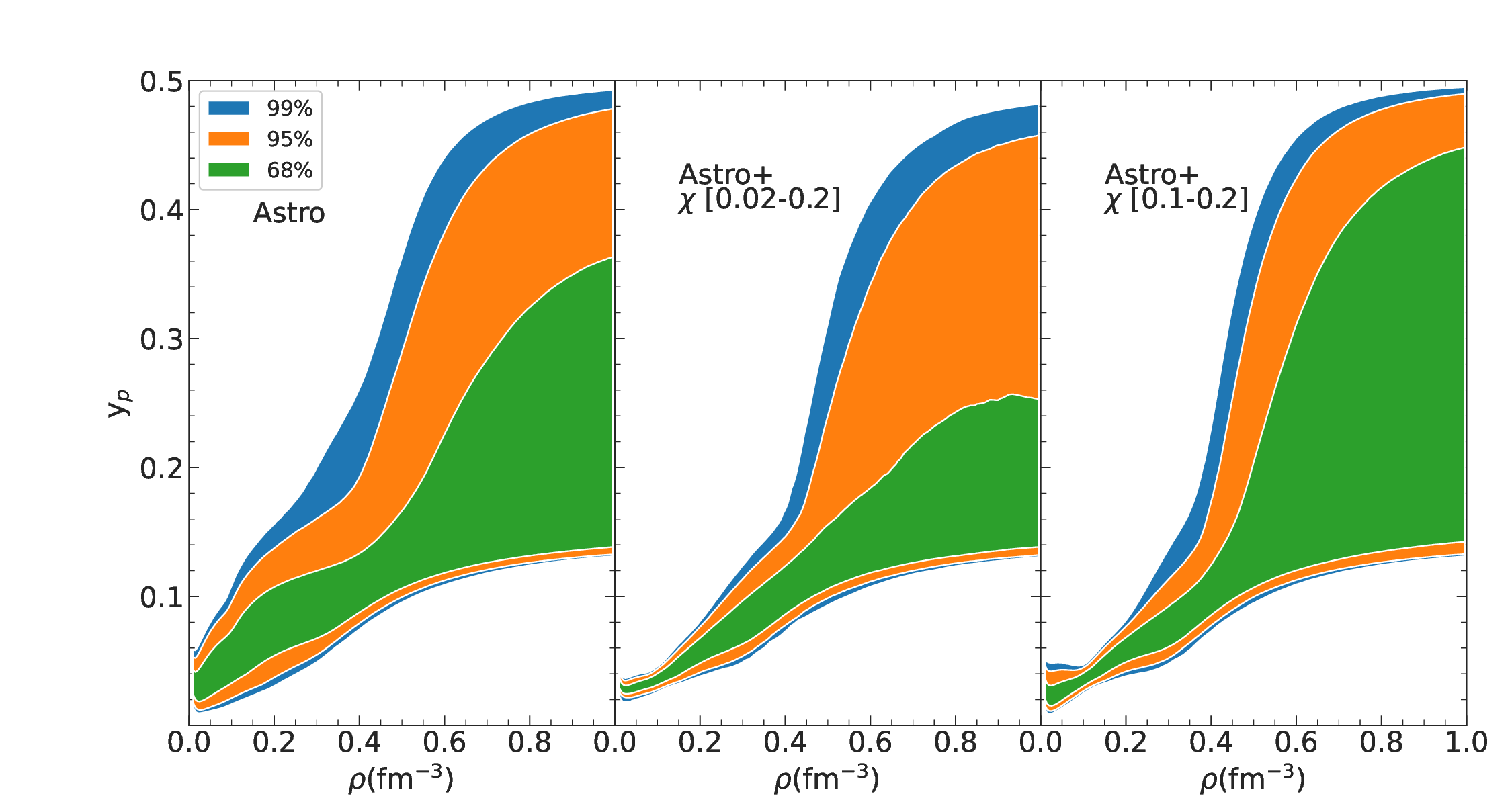}
\caption{99\%, 95\% and 68\% quantiles for the distribution of the proton fraction of beta-eq matter as a function of the baryonic density for the astrophysical posterior (left), the posterior with chiral constraint applied from $0.02$fm$^{-3}$ to $0.2$fm$^{-3}$ (center) and the posterior with chiral constraint applied from $0.1$fm$^{-3}$ to $0.2$fm$^{-3}$ (right). } 
    \label{Fig_yp_full}
\end{figure*}

\begin{figure*}
   \centering
  \hskip-2.0cm
  \includegraphics[width=1.1\linewidth,angle=0]{ 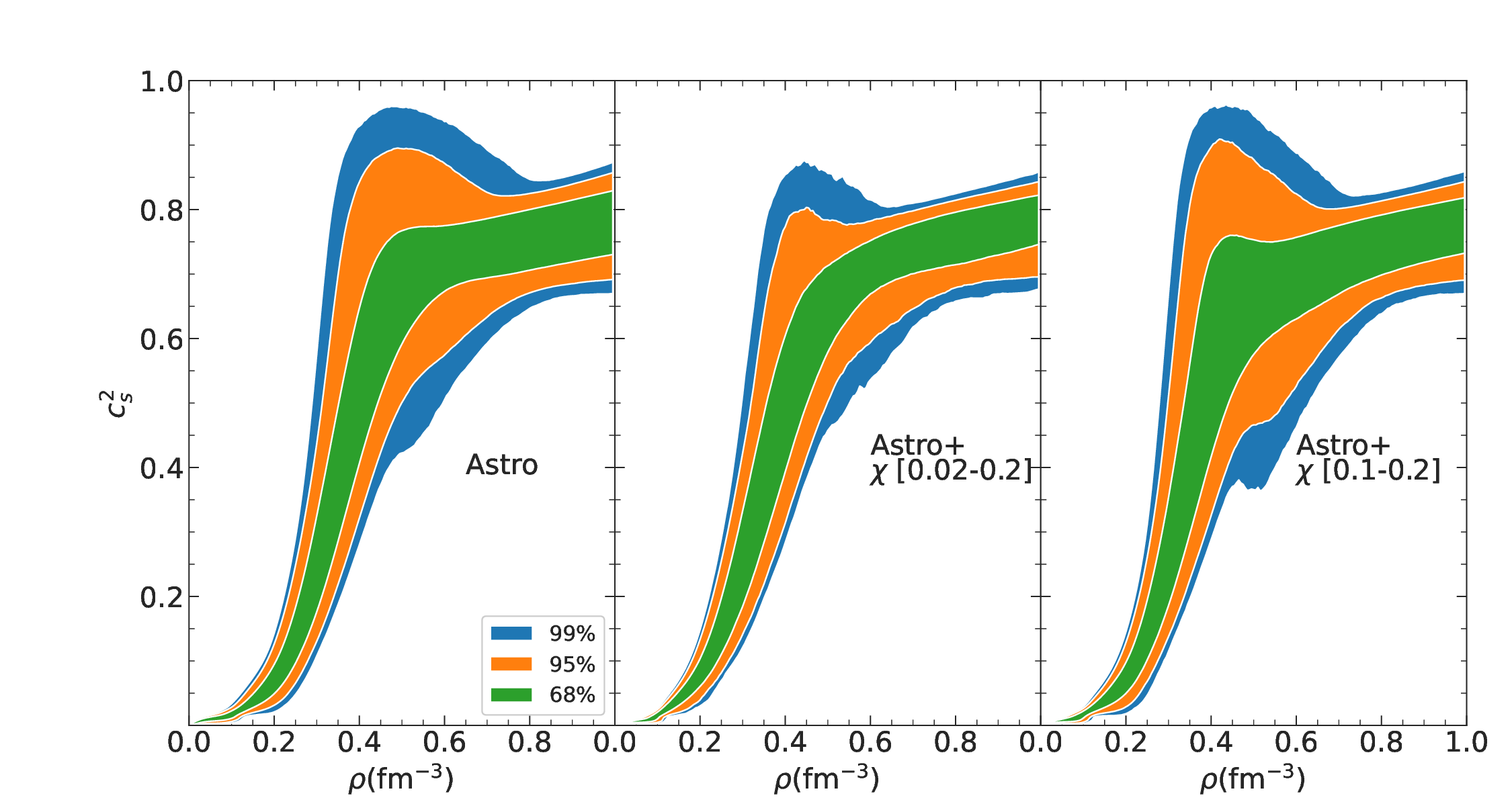}
\caption{99\%, 95\% and 68\% quantiles for the distribution of the speed of sound squared as a function of the baryonic density for the astrophysical posterior (left), the posterior with chiral constraint applied from $0.02$fm$^{-3}$ to $0.2$fm$^{-3}$ (center) and the posterior with chiral constraint applied from $0.1$fm$^{-3}$ to $0.2$fm$^{-3}$ (right)}
    \label{Fig_cs_full}
\end{figure*}

We now proceed in our analysis by studying how changing the range in which we apply the chiral constrain changes the posterior distribution. 
 It was reported in previous studies that the constraints from ab-initio nuclear theory have a much smaller impact on the determination of astrophysical quantities if they can only be trusted up to densities of the order of the saturation density or less \cite{Tews_2018,Tews:2018kmu}. The argument there is that the deconfinement transition density is not theoretically known, and in the case of a very early phase transition the nucleonic $\chi$-EFT theory would break down. Here, we make the explicit hypothesis of nucleonic degrees of freedom at any density, therefore we do not question the validity of the chiral band up to the break-down density of the chiral expansion. 
 
 However, it was remarked in ref.\cite{DinhThi2023,DinhThi2023a} that the very low density behavior of neutron matter, that is best constrained by the ab-initio theory, can have an important influence on astrophysical variables, notably the ones concerning the NS crust.
For this reason, in Fig.\ref{Fig_yp_full} and Fig.\ref{Fig_cs_full} we show the same distributions as in Fig.\ref{Fig_yp_filt} and Fig.\ref{Fig_cs_filt}, but comparing the astrophysical posterior (left panels) with a distribution 
additionally conditioned by the chiral constraint applied from $\rho=0.02$ fm$^{-3}$ to $\rho=0.2$ fm$^{-3}$ { (middle panels)}, comprising the region of the inner crust and of the outer core, and a second posterior 
in which we apply the constraint only from $\rho=0.1$ fm$^{-3}$ to $\rho=0.2$ fm $^{-3}$ { (right panels)}, that is approximately only in the outer core of the star. 

In the case of the proton fraction, we can see from   Fig.\ref{Fig_yp_full} that, when applying the constraint on the whole density range, the composition of beta equilibrated matter is well constrained also at very high densities, while its effect is limited to densities up to $\approx 3\rho_{sat}$, if the crust region is excluded { (right panel)}. 
A similar effect, though less pronounced, is also seen on 
 the speed of sound squared displayed in Fig.\ref{Fig_cs_full}:  the low density part of the chiral constraint  has the role of  considerably narrowing the sound speed squared marginalized posterior at intermediate densities ($\approx 2-4\rho_{sat}$).
\\
This surprising result can be understood if we consider the energy functional in terms of the Taylor expansion around saturation governed by the different NMPs. In these terms, the density range $[0.1-0.2]$ covers a relatively small interval around the saturation density $\rho_{sat}\approx 0.15$ fm$^{-3}$. In this region, a constraint on the energy density essentially pins down the values of the first and second order NMPs, namely $E_{sat}$, $K_{sat}$, $J_{sym}$, $L_{sym}$, $K_{sym}$, leaving the high order parameters $Q_{sat,sym}$ and $Z_{sat,sym}$ essentially unconstrained. Those parameters play the leading role in the functional behavior above $\approx 3-4\rho_{sat}$, and this is why the distributions are not considerably narrowed at very high densities. 
However, the same high order parameters $Q_{sat,sym}$ and $Z_{sat,sym}$ can have also a sizeable impact on the very low density regime $\rho<0.1$ fm$^{-3}$, if the uncertainty on the energy density is very small, as it is the case of the ab-initio calculations. Since we use the same energy density functional in the crust and core region (compare Eq.(\ref{En_dens_hom}) and Eq.(\ref{En_dens})), the uncertainty reduction issued from the low density constraint thus propagates to the highest densities. This was not seen in previous analyses \cite{universe7100373,Char_2023,Mondal:2022cva} because in those works the authors treated the third- and fourth order NMPs below and above the saturation independently, with the explicit purpose of avoiding any fictitious correlations imparted by high-density and low density data on the NMPs. 
\\
It is clear that this propagation of information from low to high density is physically reliable if and only if our effective Lagrangian corresponds to a complete description of nuclear matter in the whole density regime. Beyond mean-field correlations at low density, and an additional complexity of the couplings or the contribution of hyperons might well destroy this correlation and effectively decouple the two regimes. In this sense, we can consider the results obtained applying the chiral constraint in the whole density range as more precise, but dependent on the specific lagrangian formulation of the meta-model, while the use of the chiral constraint only around saturation leads to more general results, only conditioned by the nucleonic hypothesis.

\begin{figure*}
    \centering

       \includegraphics[width=0.8\linewidth,angle=0]{ 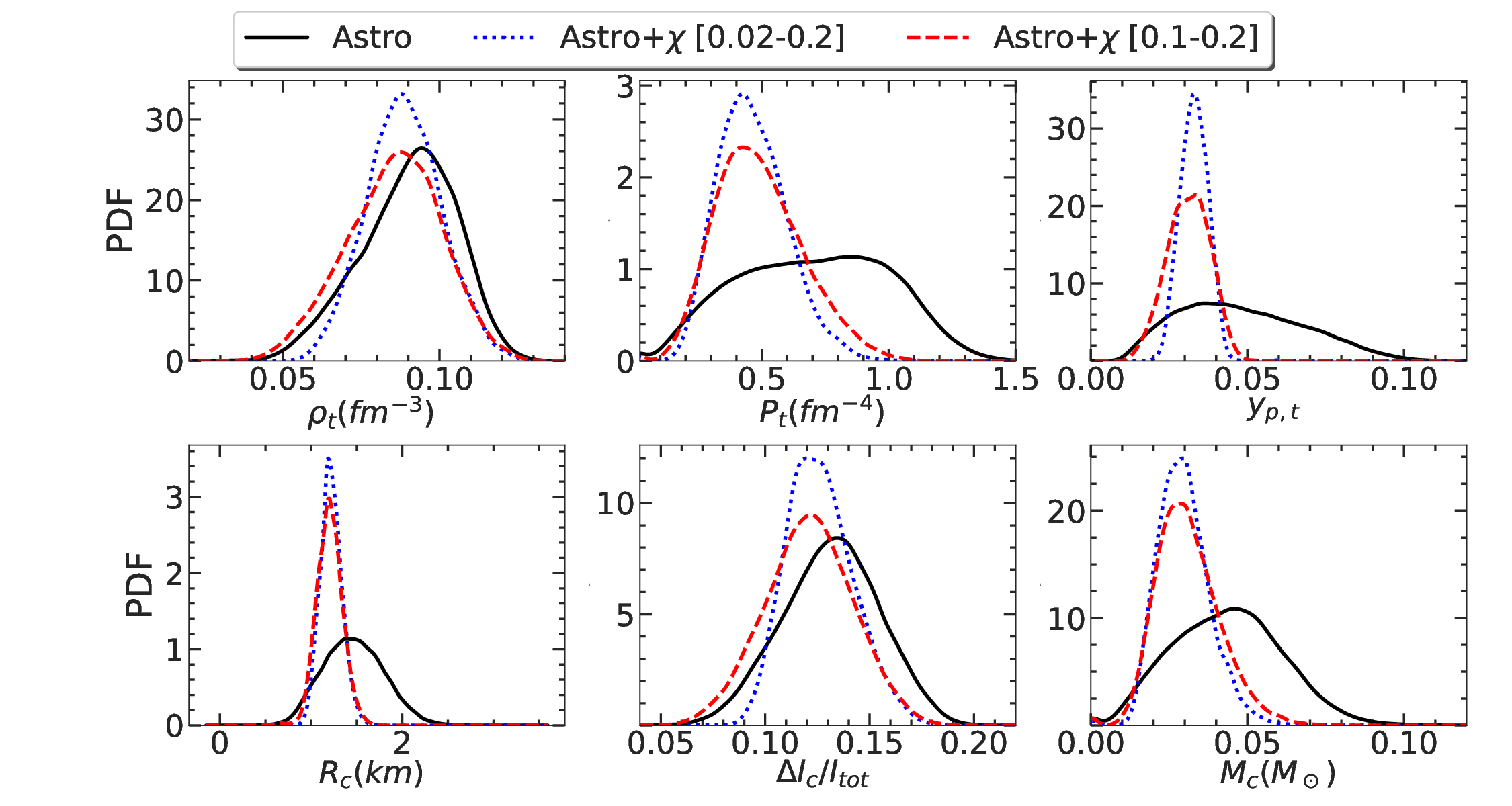}   
    \caption{PDF of the crust-core transition density (top left), pressure (top center) and proton fraction (top right), as well as of the crustal radius (bottom left), moment of inertia (bottom center) and mass (bottom right) of a $1.4M_\odot$ NS for the astrophysical posterior distribution (solid black), the posterior with chiral constraint applied from $0.02$fm$^{-3}$ to $0.2$fm$^{-3}$ (dotted blues) and the posterior with chiral constraint applied from $0.1$fm$^{-3}$ to $0.2$fm$^{-3}$ (dashed red)}.
    \label{Fig_crust_chiral}
\end{figure*}

In Fig.\ref{Fig_crust_chiral} we show the effect of the chiral constraint on the properties of the crust. In this case, the application of the constraint also in the low density region is indubitably the most realistic choice\cite{DinhThi2023}, and it systematically leads to a narrowing of the distributions as expected. The residual uncertainty is linked to the fact that we also consider the uncertainty on the nuclear surface properties, together with the ones on the bulk behavior, see Section \ref{sec:crust}. Interestingly, the crustal properties obtained using the chiral constraint over the whole density range are in very good agreement with the non-relativistic meta-model results of ref.\cite{DinhThi2023}, which is based on a completely different ansatz for the energy functional. This shows that our results, properly conditioned with the available information from ab-initio theory, can be considered as general predictions only dependent on the hypothesis of purely nucleonic degrees of freedom. 

We also see that, while the crust-core transition density and the moment of inertia of the crust are not strongly affected by the constraint, all the other distributions become much more peaked when we apply the constraint. In particular the distributions are shifted to lower values for all quantities, preferring thinner crusts and a crust-core transition at lower pressures and proton fractions.

\begin{table*}
    \centering
    \hskip-2.5cm
    \begin{tabular}{cccc|ccc|ccc}
    \hline
    \hline 
    &  \multicolumn{3}{c}{Astro}&\multicolumn{3}{c}{Astro+$\chi$}& \multicolumn{3}{c}{Astro+$\chi$+Exp}\\

    & \multicolumn{1}{c}{} & \multicolumn{2}{c}{90\% CI}& \multicolumn{1}{c}{} & \multicolumn{2}{c}{90\% CI}& \multicolumn{1}{c}{} &\multicolumn{2}{c}{90\% CI}\\
    \cline{3-4} 
    \cline{6-7} 
    \cline{9-10} 
    & \multicolumn{1}{c}{Mean} & \multicolumn{1}{c}{Min} & \multicolumn{1}{c}{Max} & \multicolumn{1}{c}{Mean} & \multicolumn{1}{c}{Min} & \multicolumn{1}{c}{Max}& \multicolumn{1}{c}{Mean} & \multicolumn{1}{c}{Min} & \multicolumn{1}{c}{Max}\\
    \hline
      $\rho_{sat}$(fm$^{-3})$   & 0.160 & 0.145 & 0.169 & 0.161 & 0.146 & 0.169 & 0.157 & 0.149 & 0.166  \\
      $E_{sat}(MeV)$    & -15.5 & -16.8 & -14.3 & -15.4 & -16.8 & -14.3 & -15.8 & -16.3 & -15.3 \\ 
      $K_{sat}(MeV)$    & 233.8 & 157.9 & 328.2 & 234.6 & 158.0 & 327.3 & 230.1 & 200.6 & 3265.6 \\ 
      $Q_{sat}(MeV)$    & -349.4 & -1302.1 & 892.4 & -350.3 & -1349.9 & 978.1 & -119.2 & -887.1 & 685.4\\ 
      $Z_{sat}(MeV)$    & 4200.7 & -8408.2  & 18443.1 & 4670.4 & -8534.8  & 18816.0 & 8241.3 & -2846.8  & 17801.4\\ 
      $E_{sym}(MeV)$    & 36.3 & 26.6 & 47.1 & 30.8 & 27.4 & 34.3 & 31.6 & 29.5 & 33.7 \\ 
      $L_{sym}(MeV)$    & 55.6 & 41.8 & 68.5 & 49.2 & 38.1 & 60.4 & 49.1 & 40.1 & 58.5 \\ 
      $K_{sym}(MeV)$    & -151.7 & -289.0 & 48.8 & -70.4 & -227.7 & 51.8 & -95.1 & -231.0 & 44.7 \\ 
      $Q_{sym}(MeV)$    & 723.7 & 44.5 & 1373.0 & 703.4 & 244.3 & 1215.6 & 707.5 & 293.2 & 1162.4 \\ 
      $Z_{sym}(MeV)$    & -4279.9 & -12023.5 & 4574.8 & -6112.0 & -11791.8 & 3500.3 & -5742.4 & -12204.6 & 3154.6 \\
      $K_{\tau}^*(MeV)$    & -485.0 & -690.0 & -277.4 & -365.9 & -501.0 & -252.5 & -389.9 & -519.4 & -288.4 \\ 
      $K_{\tau}(MeV)$    & -389.3 & -776.5 & -32.1 & -280.9 & -608.3 & 27.7 & -367.1 & -516.8 & -220.0 \\
    \end{tabular}
    \caption{Mean and extremes of the 90\% quantile for the nuclear matter parameters, as well as for $K_{\tau}^*=K_{sym}-6L_{sym}$ and for $K_{\tau}=K_{sym}-6L_{sym}-Q_{sat}L_{sym}/K_{sat} $. We show the values for the astrophysical posterior, for the posterior in which we only apply the chiral constraint and for the posterior in which we also apply the constraint coming from nuclear experiments.}
   \label{Tab_NMP_Nucl}
\end{table*}

\begin{table*}
    \hskip-2.0cm
    \centering

    \begin{tabular}{cccc|ccc|ccc}
    \hline
    \hline 
    &  \multicolumn{3}{c}{Astro}&\multicolumn{3}{c}{Astro+$\chi$}& \multicolumn{3}{c}{Astro+$\chi$+Exp}\\

    & \multicolumn{1}{c}{} & \multicolumn{2}{c}{90\% CI}& \multicolumn{1}{c}{} & \multicolumn{2}{c}{90\% CI}& \multicolumn{1}{c}{} &\multicolumn{2}{c}{90\% CI}\\
    \cline{3-4} 
    \cline{6-7} 
    \cline{9-10} 
    & \multicolumn{1}{c}{Mean} & \multicolumn{1}{c}{Min} & \multicolumn{1}{c}{Max} & \multicolumn{1}{c}{Mean} & \multicolumn{1}{c}{Min} & \multicolumn{1}{c}{Max}& \multicolumn{1}{c}{Mean} & \multicolumn{1}{c}{Min} & \multicolumn{1}{c}{Max}\\
    \hline
      $R_{1.4M_\odot}(km)$  & 12.8 & 12.0 & 13.6 & 12.7 & 12.0 & 13.4 & 12.9 & 12.3 & 13.5   \\
      $\Lambda_{1.4M_\odot}$   & 771 & 506 & 1147 & 813 & 537 & 1175 & 916 & 657 & 1222 \\ 
      $\rho^{c}_{1.4M_\odot}$(fm$^{-3})$   & 0.39 & 0.32 & 0.47 & 0.39 & 0.32 & 0.46 & 0.36 & 0.31 & 0.42 \\
      $R_{2.0M_\odot}(km)$   &  12.8 & 11.9 & 13.7 & 12.7 & 11.8 & 13.6 & 13.0 & 12.3 & 13.7 \\ 
      $\Lambda_{2.0M_\odot}$   & 83 & 44 & 137 & 86 & 45 & 141 & 104 & 61 & 148 \\ 
      $\rho^{c}_{2.0M_\odot}$(fm$^{-3})$   & 0.49 & 0.39 & 0.62 & 0.50 & 0.39 & 0.63 & 0.45 & 0.38 & 0.56 \\
      $M_{Max}(M_\odot)$   & 2.49 & 2.28  & 2.72 & 2.48 & 2.27  & 2.69 & 2.56 & 2.36  & 2.73 \\ 
      $\rho^{c}_{M_{Max}}$(fm$^{-3})$   & 0.85 & 0.71 & 0.99 & 0.85 & 0.72 & 0.99 & 0.80 & 0.71 & 0.93 \\ 
    \end{tabular}
    \caption{Mean and extremes of the 90\% quantile for the radius, tidal deformability  and central density of a $1.4M_\odot$ and $2.0M_\odot$ NS, as well as the maximum mass and the central density of the most massive star. We show the values for the astrophysical posterior, for the posterior in which we only apply the chiral constraint and for the posterior in which we also apply the constraint coming from nuclear experiments. We underline that the properties of the $1.4M_\odot$ and $2.0M_\odot$ NS are calculated only on those models that reach the respective mass.}
   \label{Tab_astro_nucl}
\end{table*}

\subsection{Experimental Nuclear Constraint}

\begin{figure*}
  \begin{minipage}[b]{0.49\linewidth}
  \centering
  \includegraphics[width=1.15\linewidth,angle=0]{ 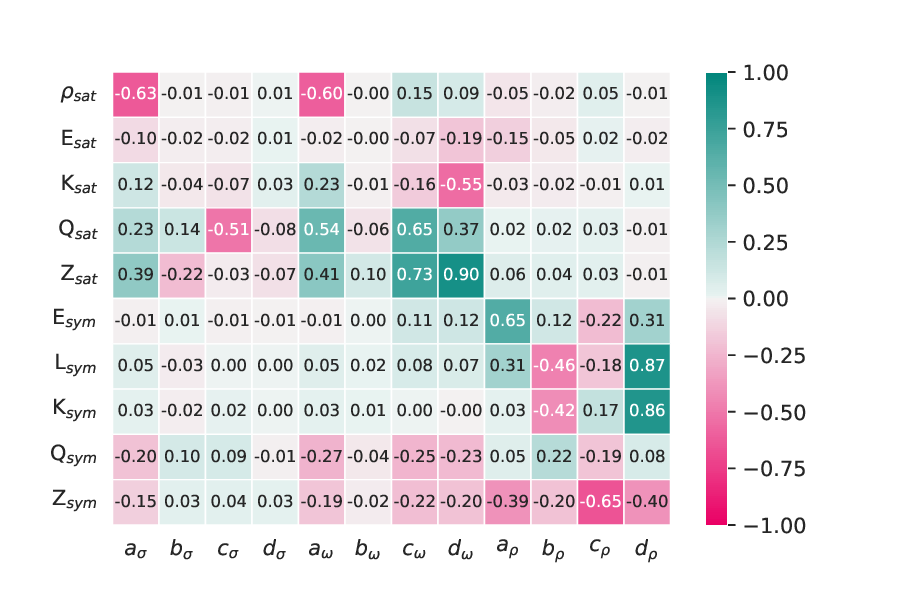}  
  \end{minipage}
  \hfill
  \begin{minipage}[b]{0.49\linewidth}
  \centering
  \includegraphics[width=1.15\linewidth,angle=0,right]{ 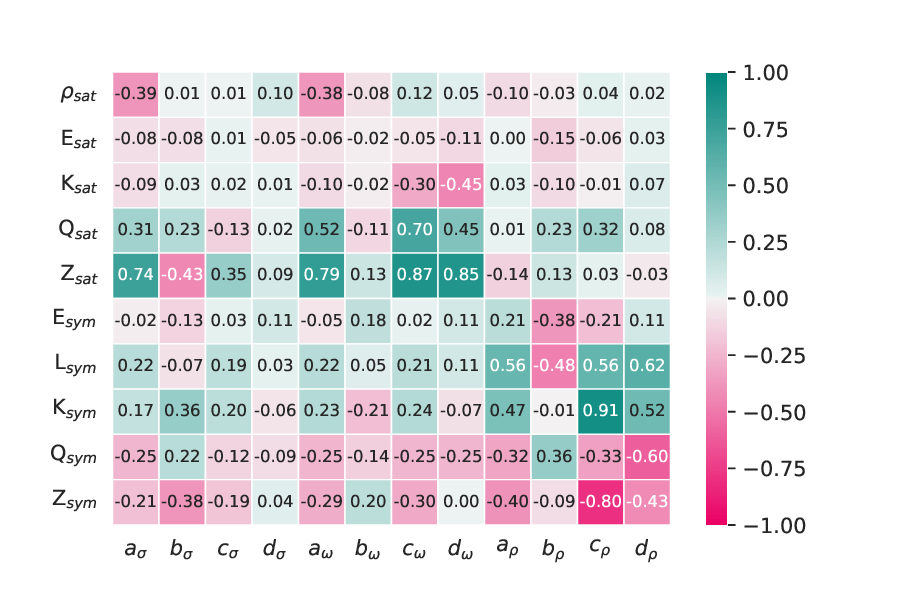}  \end{minipage}
\caption{Linear correlation between the NMPs and the 12 coupling parameters in the case of the prior (left) and  in the case of the full posterior, conditioned by all the likelihoods presented in this paper (right).}
    \label{Fig_corr_NMP_coup}
\end{figure*}

After studying the effect of the chiral constraint, we focus on the effect of the constraints on the properties of dense matter coming from nuclear experiments. In Table.\ref{Tab_NMP_Nucl} we show the comparison between mean and the extreme of the 90\% CI for the NMPs in the astrophysical posterior, a posterior distribution additionally conditioned by the chiral constraint and a second posterior obtained applying all the low density constraints (chiral + experimental). From the table we can see that the constraint does not affect only the distributions of the parameters that are directly constrained. In particular we notice that the distributions of the higher order saturation parameters $Q_{sat}$ and $Z_{sat}$ are strongly affected.  
This can be understood looking at Fig.\ref{Fig_corr_NMP_coup}, where we show the linear correlation coefficients between the NMPs and the meson coupling parameters, both in the case of the prior distribution and of the full posterior.
We can see that different NMPs appear to be correlated to the same coupling parameters. The most prominent example is given by $\rho_{sat}$, $Q_{sat}$ and $Z_{sat}$, which are all mainly correlated to $a_\sigma$ and $a_\omega$. This explains why, constraining some NMPs, we can see a strong effect also on the others. Moreover, this explains the correlation observed previously between $\rho_{sat}$ and the astrophysical observables of NS, since the observables are strongly correlated to the higher order saturation NMPs.

\begin{figure*}
    \centering
       \includegraphics[width=0.8\linewidth,angle=0]{ 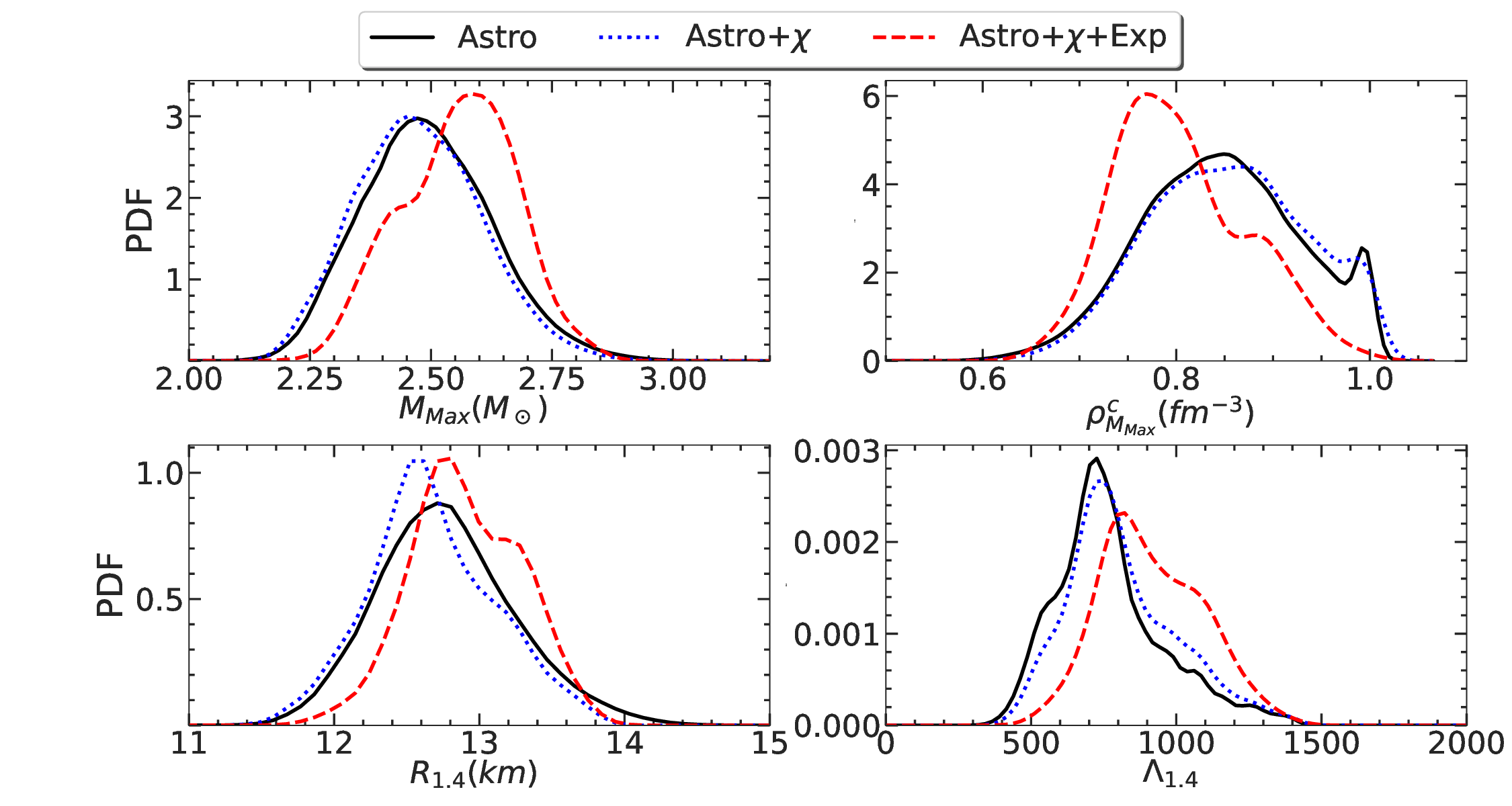}   
    \caption{PDF of the maximum mass (top left), the central density of the most massive star (top right) and of the radius (bottom left) and tidal deformability (bottom right) of a $1.4M_\odot$ NS for the prior distribution (solid black), the posterior in which we only apply the chiral constraint (dotted blue) and the posterior in which we also apply the constraint coming from nuclear experiments (dashed red).}
    \label{Fig_atro_nucl}
\end{figure*}

Finally, from the table it can also be seen that all the applied constraints give a distribution of $K_{sym}$ centered on small negative values. This agrees with the results of \cite{Reinhard:2022inh,Mondal:2022cva}, and appears to be in contrast with the solution given in \cite{Reed:2023cap} to conciliate the results of PREX and CREX.\\
In Fig.\ref{Fig_atro_nucl} we show the effect of both the chiral constraint and the nuclear experimental constraint on the same selected  observables of NS already showed in Fig.\ref{Fig_astro_astro}). Here it can be seen how the nuclear experimental constraints have an important effect on the distribution of both the maximum mass and the central density of the most massive star.

In particular we see that the distribution of the maximum mass is shifted to higher values, making our results compatible with the hypothesis of the second object in GW190814 \cite{Abbott_2020} being a NS. For what concerns the central density, on the other hand, we see that, after applying the experimental filter, the distribution is shifted to lower values, with a peak around $0.8$fm$^{-3}$. 

\begin{figure*}
  \begin{minipage}[b]{0.49\linewidth}
  \centering
  \includegraphics[width=1.15\linewidth,angle=0]{ 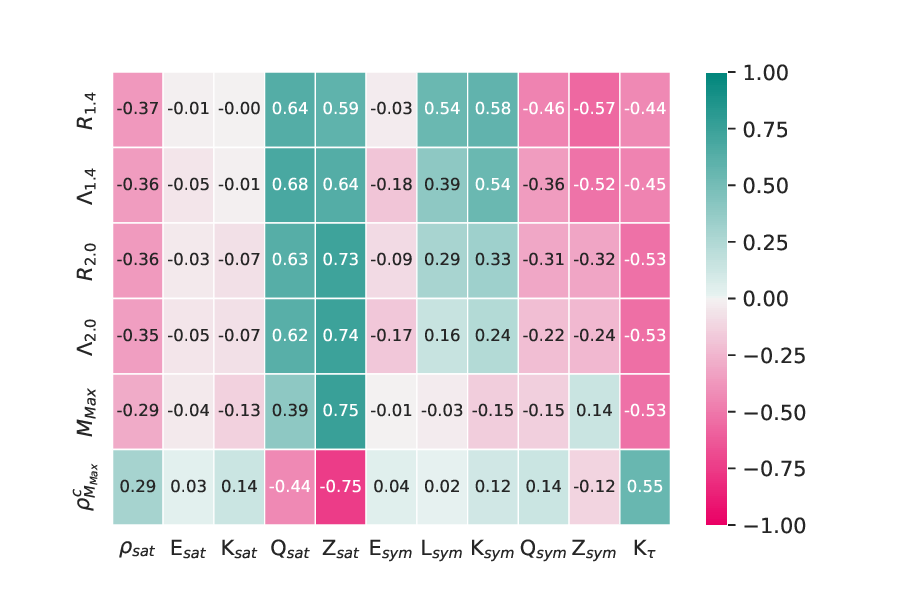}  
  \end{minipage}
  \hfill
  \begin{minipage}[b]{0.49\linewidth}
  \centering
  \includegraphics[width=1.15\linewidth,angle=0,right]{ 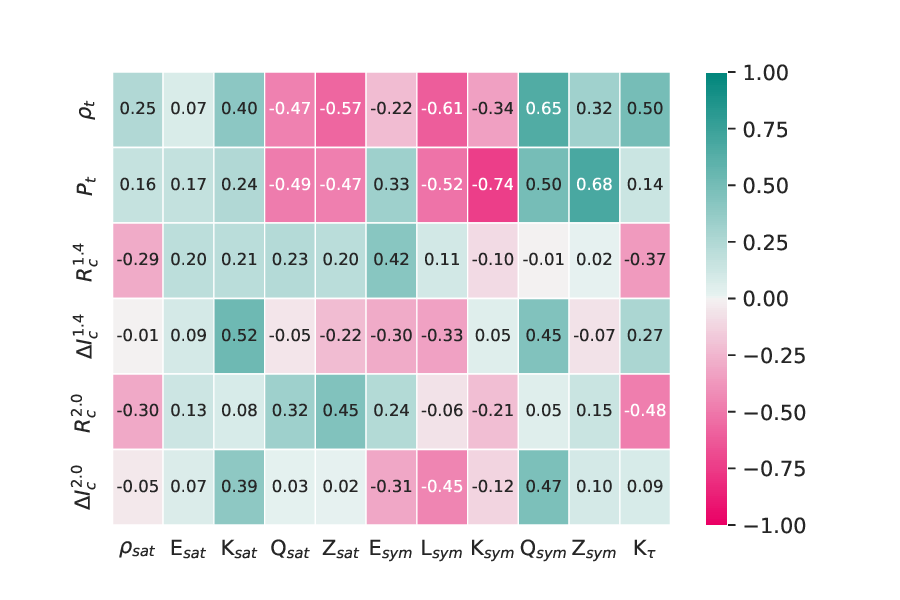}  \end{minipage}
\caption{Linear correlation between the NMPs and the astrophysical observables of NS (left) and the crustal properties of NS (right), in the case of the full posterior, conditioned by all the likelihoods presented in this paper.}
    \label{Fig_astro_crust_corr_NMP}
\end{figure*}

\begin{figure*}
    \centering
       \includegraphics[width=0.8\linewidth,angle=0]{ 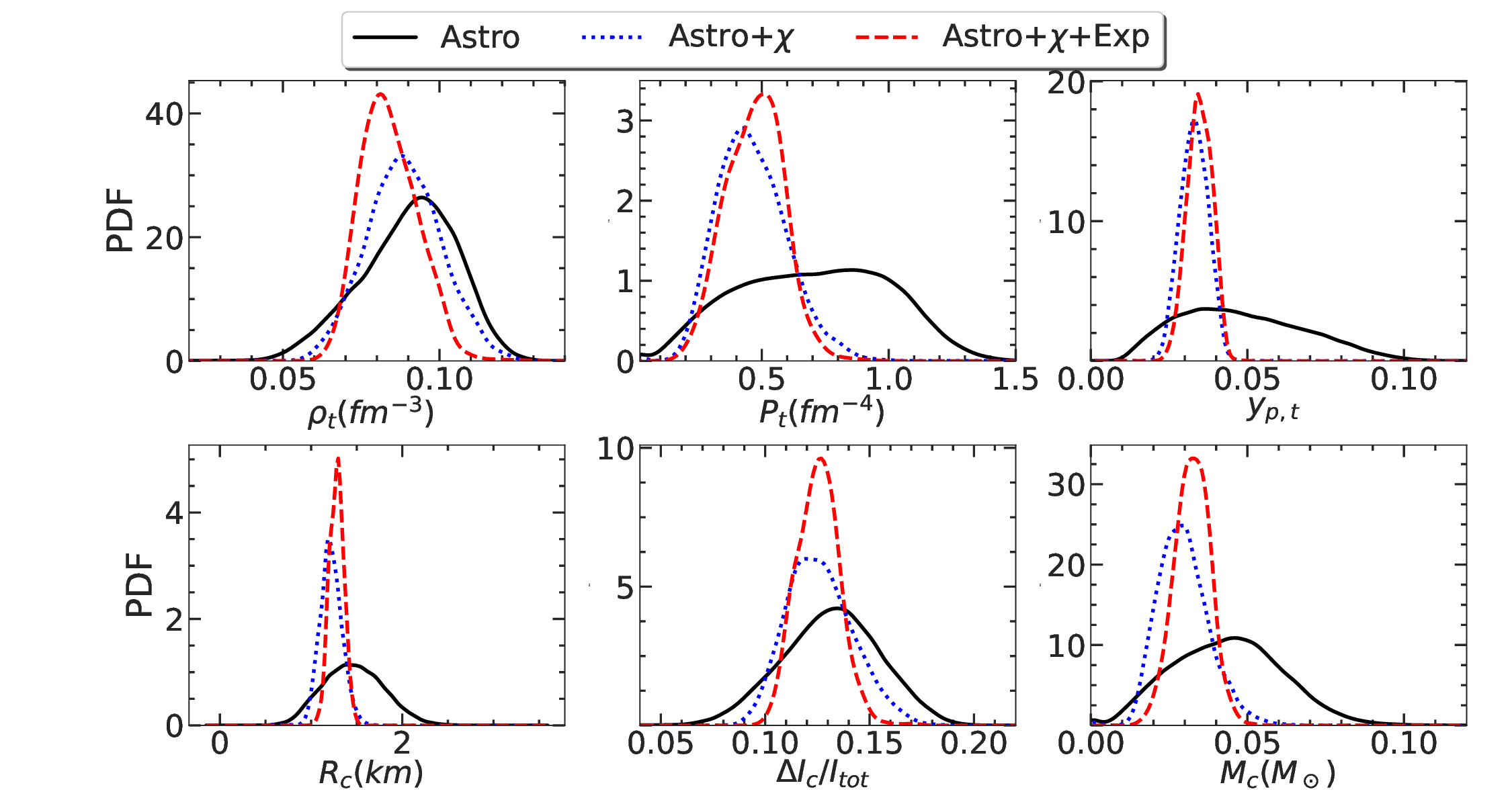}   
    \caption{PDF of the crust-core transition density (top left), pressure (top center) and proton fraction (top right), as well as of the crustal radius (bottom left), moment of inertia (bottom center) and mass (bottom right) of a $1.4M_\odot$ NS for the astrophysical posterior distribution (solid black), the posterior with astrophysical constraint + chiral constraint (dotted blues) and the full posterior, where all constraint are applied (dashed red)}
    \label{Fig_crust_nucl}
\end{figure*}

In Fig.\ref{Fig_crust_nucl} we show the effect of the constraints on the distributions of the same quantities showed in Fig.\ref{Fig_astro_astro}. We can see that the main effect on the distributions comes from the chiral constraint, while the experimental information on the low order NMPs does not shift significantly the distributions but only makes them slightly narrower. The only exception is the crust-core transition density, which is shifted to lower values. This is in agreement with previous studies, and can be understood from the fact that crustal properties are known to be well correlated with the density dependence of the symmetry energy, that is strongly constrained by the behavior of pure neutron matter.

\begin{figure*}
    \centering
       \includegraphics[width=0.8\linewidth,angle=0]{ 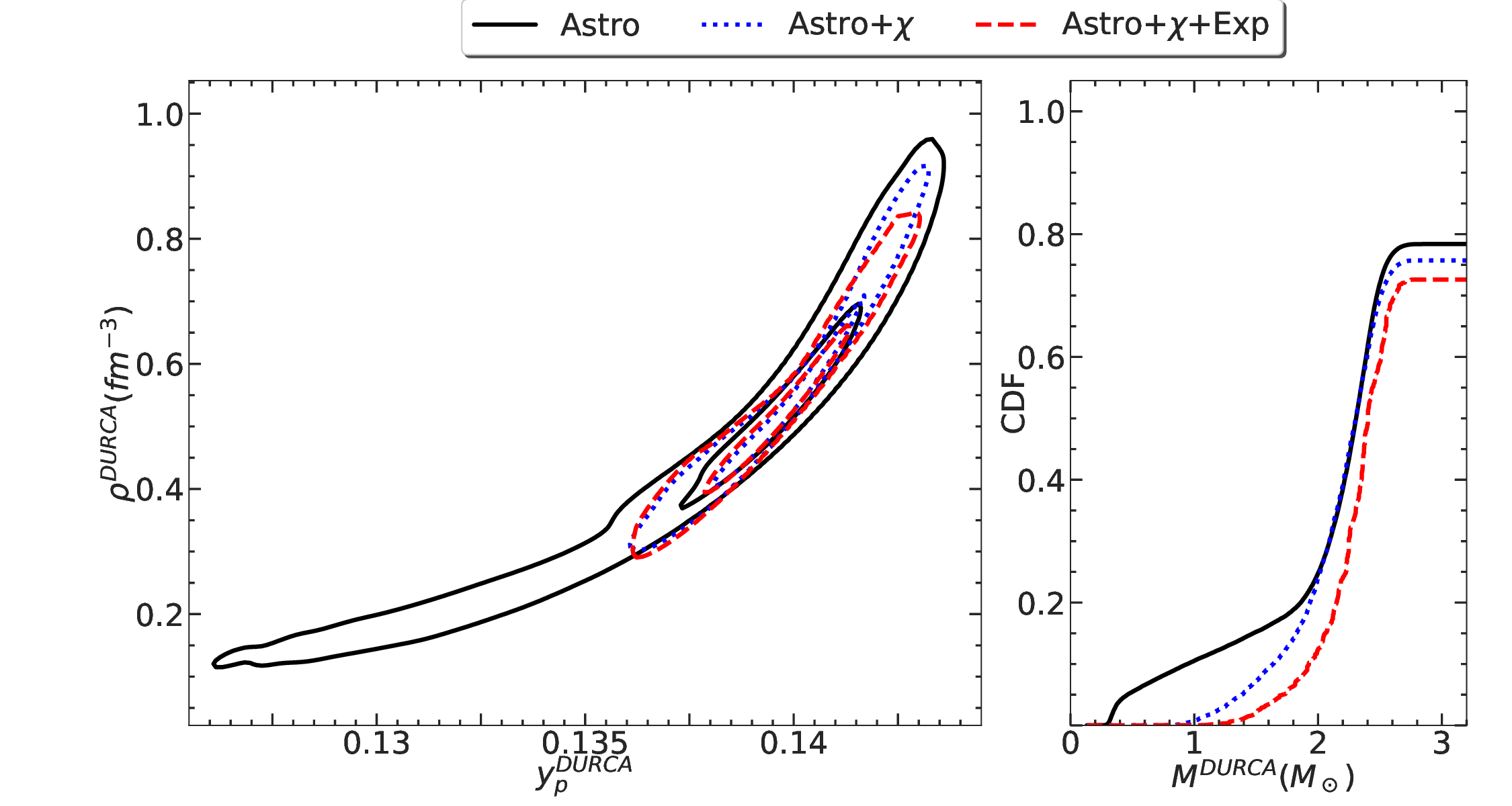}   
    \caption{Left: . correlated distribution of proton fraction and baryonic density at which the threshold for direct URCA process is reached for models allowing DURCA at densities lower than the central density of the most massive star. Right: direct URCA cumulative distribution function for a given mass. In both panels we show the astrophysical posterior distribution (solid black), the posterior with astrophysical constraint + chiral constraint (dotted blues) and the full posterior, where all constraint are applied (dashed red). 
    }
    \label{Fig_URCA}
\end{figure*}

Finally, in Fig.\ref{Fig_URCA} we show the correlation between the proton fraction and baryonic density at which the threshold for the direct URCA process is reached. The distribution is calculated taking into account only the models in which the threshold is reached before the central density of the most massive star, which amounts roughly to $83\%$ of the models in the prior. As expected, the two quantities are very well correlated in all cases, and we can see that the nuclear constraints (both theoretical and experimental) rule out the models with very low threshold density. We also show the probability of observing direct URCA process as a function of the NS mass. We  notice that, when all constraints are applied, the probability of DURCA globally decreases. It can be seen that the probability of having direct URCA process below $2 M_\odot$ is very low ($\approx 10\%$), and it becomes negligible for canonical $1.4 M_\odot$ neutron stars. This means that a potential observation of fast cooling could play the role of a smoking gun for the presence of non-nucleonic degrees of freedom.

To conclude, the results presented in this section outline the important contribution of laboratory experiments to pin down the behavior of ultra-dense matter and interpret the astrophysical observations\cite{sorensen2023dense}. However, a word of caution is in order. Indeed, nuclear physics experiments do not directly probe the NMPs, and estimations of the latter can only be obtained through model comparison. Because of that, the likelihood model of the experimental constraint employed in this work, as well as the values of the mean and standard deviation for the NMPs of Table \ref{Tab_NMP_constraint}, though coming from a compilation of different independent studies, might still be affected by some model dependence.

\subsection{Publicly Available Models}\label{sec:models}

\begin{figure*}
  \begin{minipage}{\linewidth}
  \centering
  \includegraphics[width=0.5\linewidth,angle=0]{ 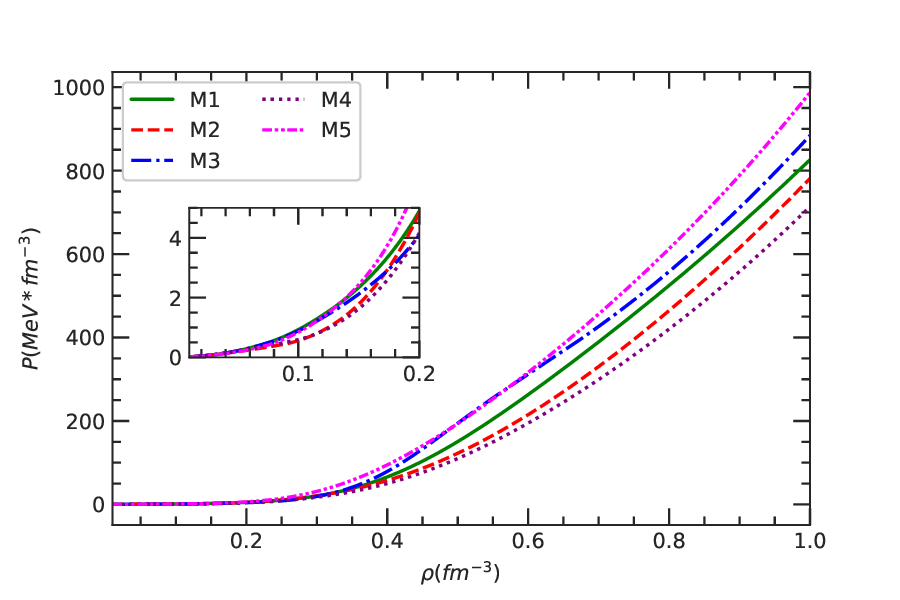}  
  \end{minipage}
  \vspace{0.5em} 
  \begin{minipage}[b]{0.49\linewidth}
  \centering
  \includegraphics[width=\linewidth,angle=0]{ 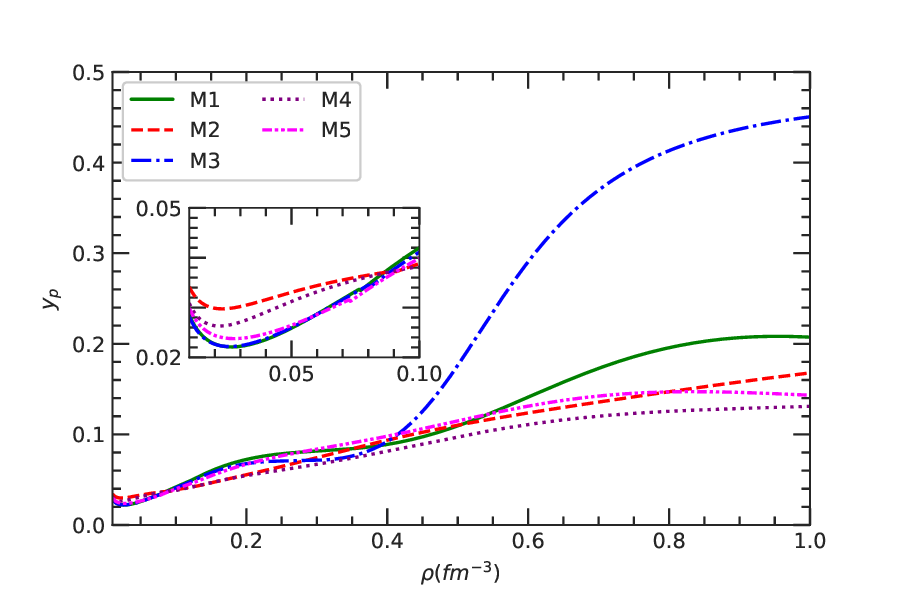}  
  \end{minipage}
  \hfill
  \begin{minipage}[b]{0.49\linewidth}
  \centering
  \includegraphics[width=\linewidth,angle=0]{ 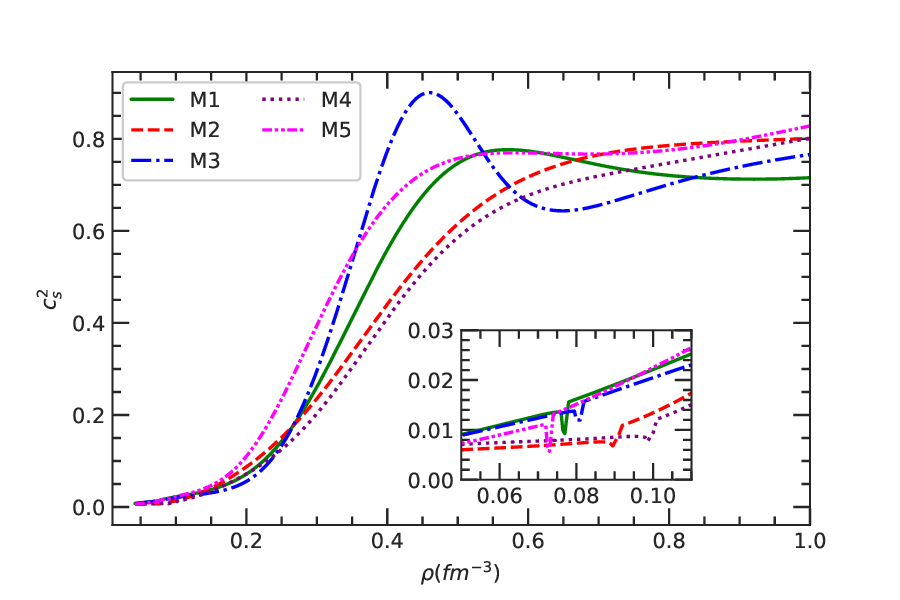}  \end{minipage}
\caption{Pressure (top), proton fraction (bottom left) and speed of sound (bottom right) as a function of density for the five selected models.}
    \label{Fig_selected_models_beta}
\end{figure*}

\begin{figure*}
  \begin{minipage}[b]{0.49\linewidth}
  \centering
  \includegraphics[width=\linewidth,angle=0]{ 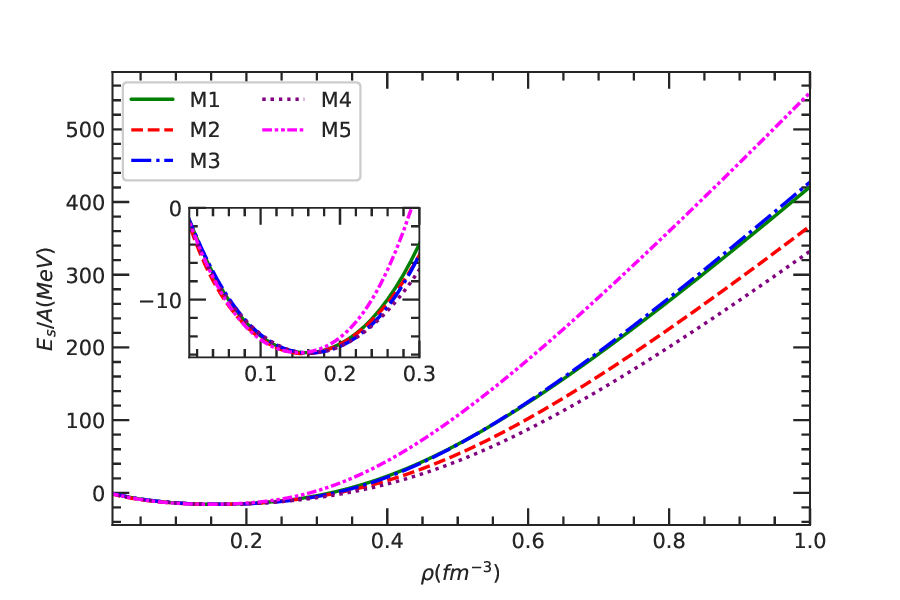}  
  \end{minipage}
  \hfill
  \begin{minipage}[b]{0.49\linewidth}
  \centering
  \includegraphics[width=\linewidth,angle=0]{ 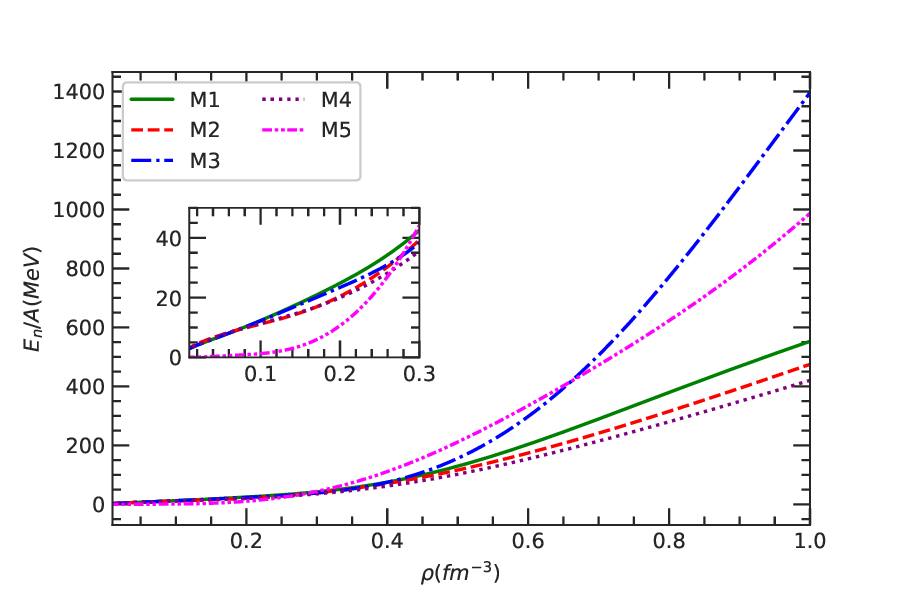}  \end{minipage}
    \begin{minipage}[b]{0.49\linewidth}
  \centering
  \includegraphics[width=\linewidth,angle=0]{ 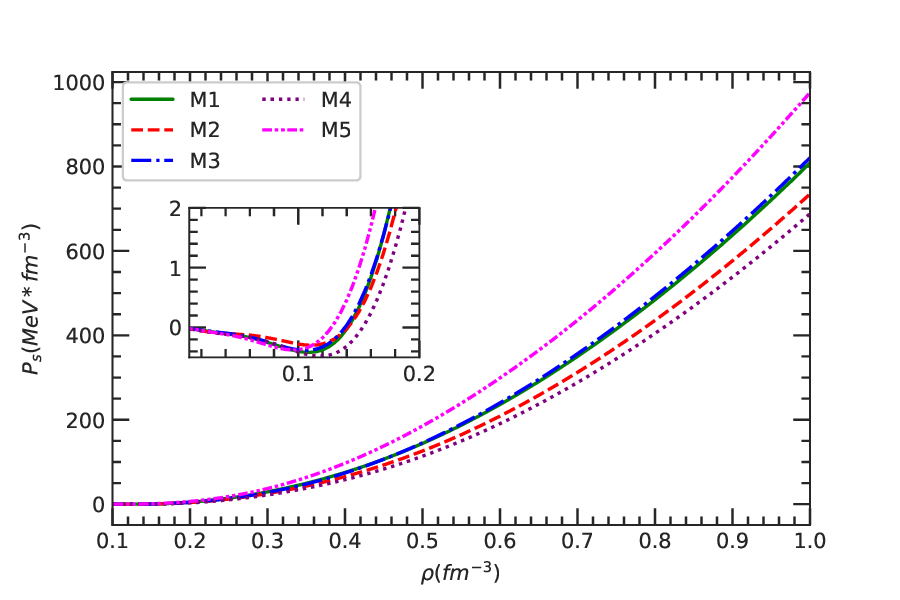}  
  \end{minipage}
  \hfill
  \begin{minipage}[b]{0.49\linewidth}
  \centering
  \includegraphics[width=\linewidth,angle=0]{ 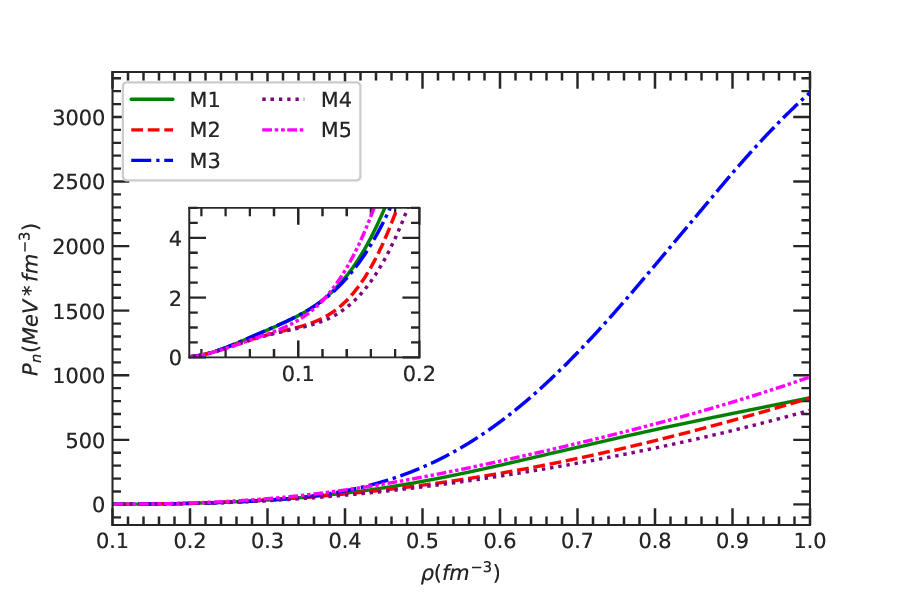}  \end{minipage}
\caption{Energy per baryon (top) and pressure (bottom) of symmetric (left) and neutron (right) matter as a function of density for the five selected models.}
    \label{Fig_selected_models_fixed}
\end{figure*}

\begin{figure*}
  \begin{minipage}{\linewidth}
  \centering
  \includegraphics[width=0.5\linewidth,angle=0]{ 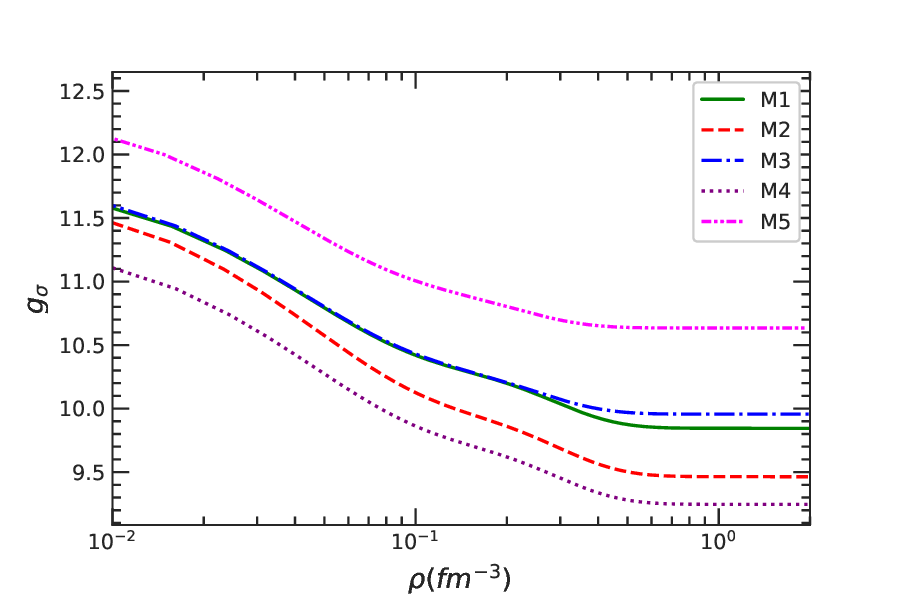}  
  \end{minipage}
  \vspace{0.5em} 
  \begin{minipage}[b]{0.49\linewidth}
  \centering
  \includegraphics[width=\linewidth,angle=0]{ 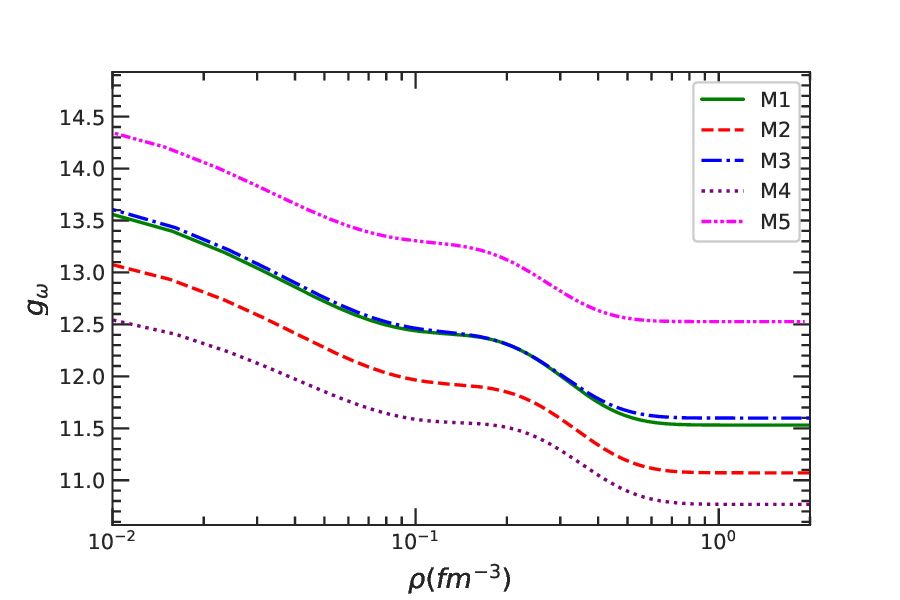}  
  \end{minipage}
  \hfill
  \begin{minipage}[b]{0.49\linewidth}
  \centering
  \includegraphics[width=\linewidth,angle=0]{ 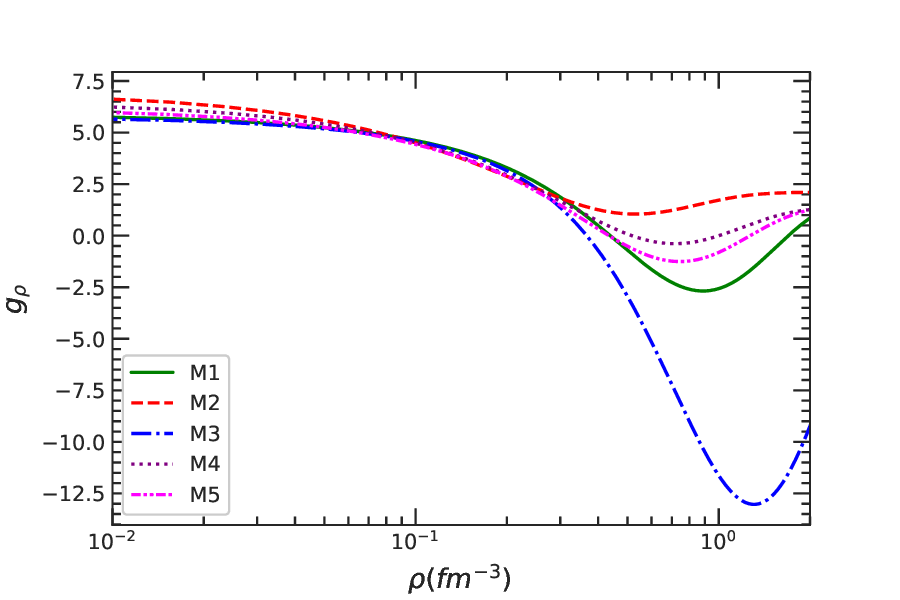}  \end{minipage}
\caption{Sigma meson (top), omega meson (bottom left) and rho meson (bottom right) couplings as a function of $\rho/\rho_0$ for the five selected models.}
    \label{Fig_selected_models_coupling}
\end{figure*}

\begin{figure*}
  \begin{minipage}[b]{0.49\linewidth}
  \centering
  \includegraphics[width=\linewidth,angle=0]{ 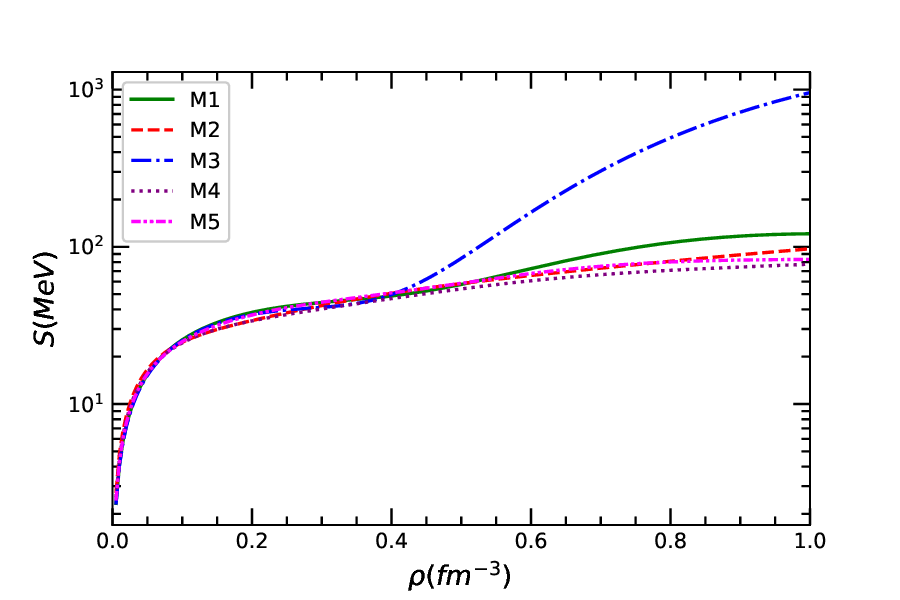}  
  \end{minipage}
  \hfill
  \begin{minipage}[b]{0.49\linewidth}
  \centering
  \includegraphics[width=\linewidth,angle=0]{ 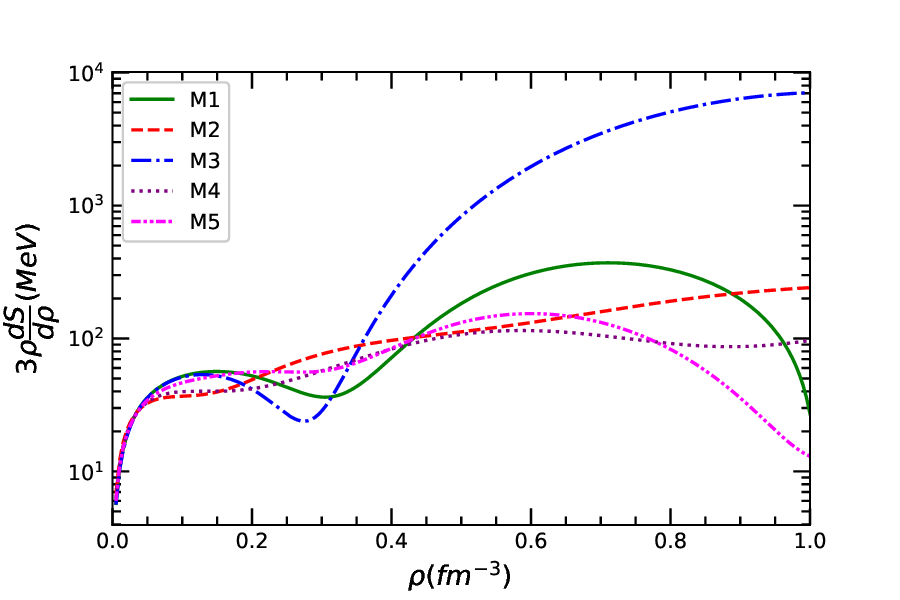}  \end{minipage}
\caption{
Symmetry energy (left) and its first derivative 
(right) as a function of density as defined in \cite{Dutra:2014qga} for the five selected models.}
    \label{Fig_selected_models_Esym}
\end{figure*}

\begin{figure*}[]
  \begin{minipage}[b]{0.49\linewidth}
  \centering
  \includegraphics[width=\linewidth,angle=0]{ 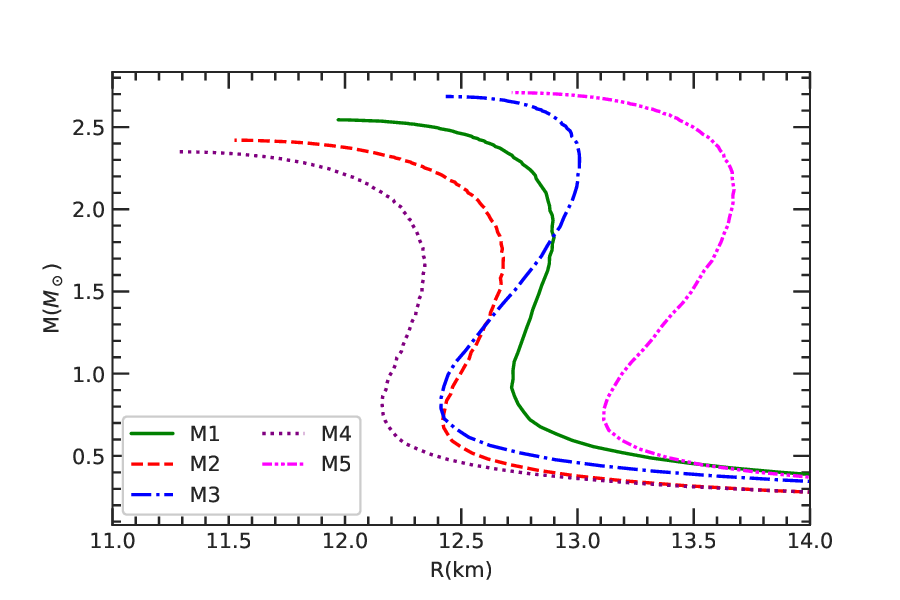}  
  \end{minipage}
  \hfill
  \begin{minipage}[b]{0.49\linewidth}
  \centering
  \includegraphics[width=\linewidth,angle=0]{ 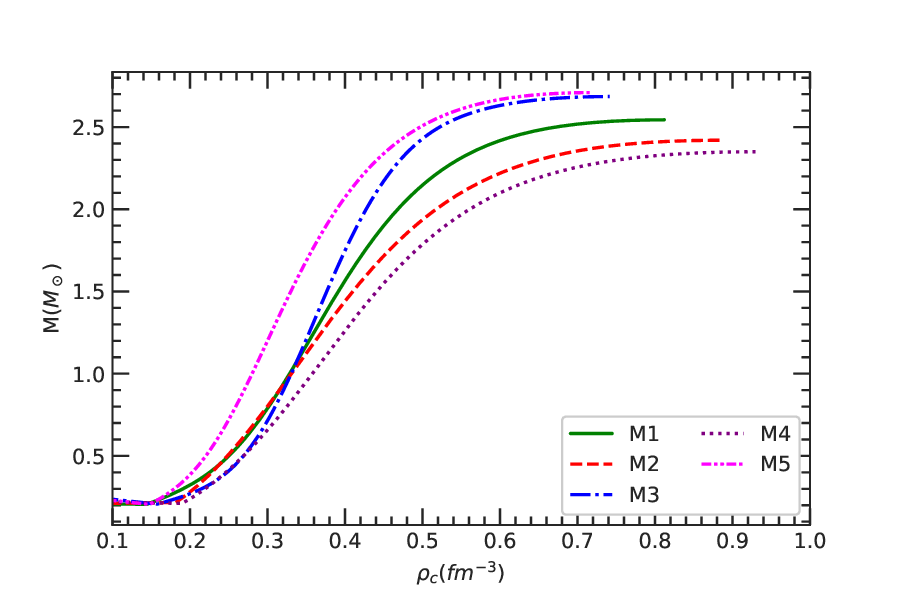}  \end{minipage}
\caption{Mass-radius relation (left) and mass as a function of central density (right) for the five selected models.}
    \label{Fig_selected_models_astro}
\end{figure*}

We end our study by selecting a subset of { five} models from our set, that are representative of the final uncertainty brought by our full posterior. These models will be made publicly available on the CompOse online EoS repository. The models are selected as follows.
The model M1 corresponds to the highest absolute likelihood in the full posterior. 
 M2 and M3 are the models corresponding to the highest weight around the extremes of the $68\%$ CI for the maximum mass; M4 and M5 are instead selected around the extremes of the $90\%$ CI for the radius of the $1.4M_\odot$ NS.

These models are selected in order to cover in a representative way the posterior distribution of masses and radii. The choice of these models also grants variety in terms of proton fraction, speed of sound and higher order NMPs.

{This can be appreciated from Fig.~\ref{Fig_selected_models_beta}, where we show the pressure (top), the proton fraction  (bottom left) and the speed of sound squared (bottom right) of $\beta-$equilibrium matter  as a function of density for the five selected models. One of the models, M5, shows a very high increase in the proton fraction around $\rho\sim 0.4$ fm$^{-3}$, a behaviour quite different from the other four models considered. This is also seen in the speed of sound squared that reaches a peak at that density, and in the neutron matter pressure, that presents a very stiff behaviour, as shown in Fig.~\ref{Fig_selected_models_fixed}, along with the energy per particle, and the pressure and energy per particle  of symmetric matter. This peculiar behavior  can be traced back to the density-dependence of the model couplings, shown in Fig.\ref{Fig_selected_models_coupling} in the vector-isovector ($\rho$),  vector-isoscalar ($\omega$), and scalar-isoscalar ($\sigma)$ channel. From this figure we can see that the $g_{\rho}$ coupling starts decreasing at the same density where the proton fraction starts increasing, saturating at $\rho\sim 1$ fm$^{-3}$, where the proton fraction reaches a maximum. The other models only have a maximum of $0.1<y_p<0.2$ because they do not have this steep decrease in the vector-isovector channel. 

In turn, the possible complex density dependence of the isovector coupling shown in Fig.\ref{Fig_selected_models_coupling} induces a non-trivial behavior of the density dependence of the symmetry energy, defined as

\begin{equation}
{\cal{S}}(\rho)=\frac 18 \frac{\partial^2(\cal{E}/\rho)}
{\partial y_p^2}\bigg|_{y_p=1/2} \, ,
\end{equation}
  which is plotted in the left panel of Fig.~\ref{Fig_selected_models_Esym}, along with its first derivative (right panel of the same figure). In this Figure, we can see 
  that the high value of the proton fraction of M3 and its steep variation with the density is perfectly correlated with the behavior of the symmetry energy. It is also interesting to see in Fig.~\ref{Fig_selected_models_beta} that such non-trivial behavior in the poorely known isospin channel can result in a well-defined peak in the sound speed. Though the possible physical meaning of a strong decrease of the effective $\rho$ coupling is not clear, this result shows that peaks in the sound speed can very well be obtained in purely nucleonic models, and they do not necessarily signal the presence of phase transitions.

{ In Fig.~\ref{Fig_selected_models_astro}, we show the mass versus radius (left) and versus central density (right) for the five models considered. The models with the highest maximum mass are the ones that present the stiffest $\beta-$equilibrium matter EoS (see Fig.~\ref{Fig_selected_models_beta}. The model with the largest central density and the smallest radius is the one that has the softest EoS, the model M4. These properties are also written in Table~\ref{Tab_astro_selected}, together with the tidal deformability $\Lambda$, and the proton fraction and correspondent density for the onset of the direct Urca processes. The model M4 is also the one that does not allow such kind of processes because the symmetry energy is very low (see Fig.~\ref{Fig_selected_models_Esym}), that favors very asymmetric matter matter (see Fig.~\ref{Fig_selected_models_beta}).}

\begin{table*}
    \centering
    \hskip-1.5cm
    \begin{tabular}{c|ccccc}
    \hline
    \hline 
    &  M1 & M2 & M3 & M4 & M5 \\
    \hline
      $R_{1.4M_\odot}(km)$  & 12.8 & 12.6 & 12.7 & 12.3 & 13.4  \\
      $\Lambda_{1.4M_\odot}$   & 811 & 789 & 792 & 657 & 1169\\ 
      $\rho^c_{1.4M_\odot}(km)$  & 0.38 & 0.39 & 0.37 & 0.42 & 0.32  \\
      $R_{2.0M_\odot}(km)$   &  12.9 & 12.6 & 13.0 & 12.2 & 13.7\\ 
      $\Lambda_{2.0M_\odot}$   & 89 & 76 & 99 & 61 & 143  \\ 
      $\rho^c_{2.0M_\odot}(km)$   &  0.47 & 0.52 & 0.43 & 0.56 & 0.39\\ 
      $M_{Max}(M_\odot)$   & 2.54 & 2.42  & 2.69 & 2.35 & 2.71\\ 
      $\rho^{c}_{M_{Max}}$(fm$^{-3})$   & 0.82 & 0.89 & 0.75 & 0.93 & 0.72  \\ 
      $\rho^{DURCA}$(fm$^{-3})$   & 0.59 & 0.73 & 0.46 & - & 0.66  \\ 
      $y_p^{DURCA}$   & 0.138 & 0.139 & 0.136 & - & 0.139  \\ 
      $M^{DURCA}(M_\odot)$   & 2.40 & 2.38 & 2.25 & - & 2.70  \\ 
    \end{tabular}
    \caption{Values of the radius, tidal deformability  and central density of a $1.4M_\odot$ and $2.0M_\odot$ NS, as well as the value of the maximum mass and the central density of the most massive star and of the density and proton fraction at which the direct URCA threshold is met, for the five selected models.}
   \label{Tab_astro_selected}
\end{table*}

In Table~\ref{Tab_NMP_selected}, we show the nuclear matter properties at saturation for the five selected models. The reader can see that the models were selected in order to cover in a representative way the uncertainty of our posterior also in this case, with in particular great variations in the high order NMPs.

\begin{table*}[]
    \centering
    \hskip-1.5cm
    \begin{tabular}{c|ccccc}
        \hline
        \hline
         & M1 & M2 & M3 & M4 & M5\\
         \hline
          $\rho_{sat}$(fm$^{-3})$ & 0.157 & 0.155 & 0.161 & 0.162 & 0.148\\
          E$_{sat}(MeV)$ & -15.8 & -15.9 & -15.8 & -15.8 & -15.8\\
          K$_{sat}(MeV)$ & 230.9 & 223.9 & 225.7 & 246.0 & 227.7\\
          Q$_{sat}(MeV)$ & -177.1 & -230.1 & -283.4 & -292.9 & 45.2\\
          Z$_{sat}(MeV)$ & 7966.3 & 2141.8 & 11174.8 & -1397.4 & 14552.1\\
          J$_{sym}(MeV)$ & 33.9 & 30.2 & 33.7 & 30.9 & 31.4\\
          L$_{sym}(MeV)$ & 56.3 & 40.2 & 50.9 & 40.1 & 52.3\\
          K$_{sym}(MeV)$ & -181.4 & -59.8 & -231.0 & -111.8 & -113.5\\
          Q$_{sym}(MeV)$ & 350.1 & 1007.2 & 322.9 & 958.5 & 539.8\\
          Z$_{sym}(MeV)$ & -1700.3 & -8829.1 & 1343.3 & -5795.1 & -5362.5\\
    \end{tabular}
    \caption{Values of the NMP for the five selected models.}
   \label{Tab_NMP_selected}
\end{table*}

\section{Conclusions} \label{Sec_fin}

In this work we provide general predictions on static properties of neutron stars (NS)
within a Bayesian framework conditioned by a large set of laboratory, astrophysical and theoretical constraints.  Our equation of state setting explores a large parameter space compatible with the requirement that both in the crust and in the core of the star the EoS is derived from a same Lagrangian density in the mean-field approximation with baryonic constituents limited to neutrons and protons. 
The general functional form is obtained including scalar-isoscalar, vector-isovector and  vector-isoscalar effective mesons. A wide exploration of the different possible behaviors of the energy at the high baryonic densities expected in the core of the most massive NS is obtained employing the phenomenological flexible expression for the density dependent couplings proposed in ref.\cite{Char_2023} and inspired by the GDFM relativistic mean-field model\cite{Gogelein:2007qa}.

The study is particularly focused on the complementary effect of different constraints on the posterior distributions of astrophysical observables.
We consider two main categories of constraints. The first family comprises the ones acting around the equilibrium density of terrestrial nuclei or below, which include the information coming from ab-initio nuclear theory employing $\chi-EFT$ interactions summarized in the compilation \cite{Huth2021}, and the estimation of five nuclear matter empirical parameters (NMPs) obtained from the analysis of different experimental nuclear data in the compilation \cite{Margueron:2017eqc}. The second class corresponds to constraints from 
astrophysical observations, which include the constraint on the maximum mass coming from the observation of J0348+0432 \cite{Antoniadis:2013pzd}, the tidal polarizability constraint coming from the observation of GW170817 \cite{TheLIGOScientific:2017qsa,Abbott_2017_2,Abbott_2019} and the constraint coming from the two combined measurements of mass and radius of NS done by NICER \cite{Riley:2019yda,Miller:2019cac,Riley:2021pdl,Miller:2021qha}.

The main results can be summarized as follows. 
First, we showed that the strong correlation between the tidal polarizability and the NS radius implies that the gravitation wave measurement of GW170817 is more effective 
in constraining the NS radius than the direct radius estimation by NICER, due to the still large systematics of the latter observations. Since this correlation depends on the EoS model, our radius estimation holds for the explicit hypothesis of nucleonic degrees of freedom, and this statement would not be correct in fully agnostic treatments of the EoS.

Concerning EoS information coming from ab-initio nuclear theory, the translation of the model dependence of the existing many-body treatments into a quantitative estimation in terms of probability is still an ongoing challenge of nuclear theory. Still, we showed that two different likelihood models, differing in the way the theoretical uncertainty band on the energy of pure neutron matter is interpreted, lead to fully compatible posterior distributions. To further tighten the predictions particularly at high density, it will be very important to assess the possible model dependence of the likelihood models concerning derived quantities such as the pressure or the sound speed.

 We also showed how, changing the density range in which the distributions are conditioned by the theoretical information, strongly affects the results. In agreement with the non-relativistic metamodelling approach of ref.\cite{DinhThi2023}, applying the constraints starting from
 densities of the order of the drip density appears crucial to get model independent estimations of crustal quantities. More surprisingly, the constraint applied in the density region corresponding to the NS crust is seen to strongly condition the behavior of 
 quantities such as the proton fraction and the speed of sound in the core of the star. This strong coupling between different density regimes is due to the treatment of the crust and the core using a unique model for all density domains.
 This is a general feature of all unified models that do not allow for extra degrees of freedom such as hyperons at high density, and are ruled by a limited number of parameters that can be precisely pinned down from low density information. 
 
 Because of that, a model dependence can appear at high density if different classes of nucleonic models, for instance those issued from RMF effective Lagrangians and those coming from Skyrme-like forces, are conditioned by the same low-density ab-initio results. However, when a looser constraint is applied limiting the conditional probability to the nuclear saturation region, fully compatible results are obtained between our relativistic meta-model and the previous similar works using different settings for the energy functional \cite{universe7100373,Mondal:2022cva,Char_2023}. This underlines again the fact that our results can be taken as general predictions under the hypothesis of a purely nucleonic composition of ultra-dense matter, and as such they can be used to test the possible presence of exotic components in the core of neutron stars.
 The only notable difference between the results presented in this paper and the ones of the non-relativistic version of the meta-model concerns the behavior of the sound speed at the highest densities and the posterior distribution of the maximum mass. The in-built causality implicit in our relativistic Lagrangian formulation leads to a controlled behavior of the sound speed without any artificial cut to exclude superluminal behaviors. This is at variance with non-relativistic functionals, where causality is often violated for masses below the maximum mass allowed by the TOV equation\cite{Margueron:2017lup}. As a consequence, a purely nucleonic content and the possible existence of very massive NS overcoming $2.5 M_\odot$ are shown to be perfectly compatible with all the present observations.   
 
  Finally, we studied the effect of the information coming from nuclear experiments, implementing gaussian likelihood models on the NMPs with means and variances as suggested by different low energy nuclear physics experiments. We showed how these constraints affect the observables of NS, notably by shifting the distribution of the central density of the most massive star to lower values. We suggested that this effect might be related to the correlation between the observables and the NMPs. It is important to stress that these constraints are always indirectly obtained by comparing experimental nuclear data to chosen models, employing different energy functionals as well as different many-body methods. For this reason, to fully assess the model dependence of the assumed likelihoods, it appears very important for the future to develop full combined Bayesian studies of the different nuclear structure observables with beyond-mean field realizations of the density functional models. 
  
  Finally, five chosen representative models that respect well all the constraints taken into account in this study, and approximately cover the residual uncertainty in our posterior distributions, are selected. This models are made publicly available and uploaded on the EoS repository CompOse, to be used in future analyses as a qualitative tool to assess the model dependence associated to the nucleonic hypothesis.

\section{Acknowledgements}
We thank Chiranjib Mondal, Prasanta Char and Micaela Oertel for useful discussions. L.S. acknowledges the PhD grant 2021.08779.BD (FCT, Portugal). Partial support from the IN2P3 Master Project NewMAC and the ANR project GWsNS, contract ANR-22-CE31-0001-01 is also acknowledged.
This work was also partially supported by national funds from FCT (Fundação para a Ciência e a Tecnologia, I.P, Portugal) under projects 
UIDB/04564/2020 and UIDP/04564/2020, with DOI identifiers 10.54499/UIDB/04564/2020 and 10.54499/UIDP/04564/2020, respectively, and the project 2022.06460.PTDC with the associated DOI identifier 10.54499/2022.06460.PTDC. HP acknowledges the grant 2022.03966.CEECIND (FCT, Portugal) with DOI identifier 10.54499/2022.03966.CEECIND/CP1714/CT0004.

\bibliographystyle{apsrev4-1}
\bibliography{refs.bib}

\end{document}